\documentclass[
reprint,
amsmath,
amssymb,
aps,
pre,
a4paper,
floatfix]{revtex4-2}

\usepackage{placeins}
\usepackage{graphicx}
\usepackage[caption=false]{subfig} 

\usepackage{dcolumn}
\usepackage{bm}
\usepackage{tikz}
\usetikzlibrary{shapes.geometric}
\newcommand\red[1]{{\color{red}#1}}
\newcommand\blue[1]{{\color{blue}#1}}
\usepackage{wasysym}

\usepackage{blindtext}
\usepackage{booktabs,multirow}
\usepackage{makecell}
\usepackage{svg}

\newcommand{\TT}{{\mathcal{T}}}
\newcommand{\av}[1]{{\mathbf a}_{#1}}
\newcommand{\bv}[1]{{\mathbf b}_{#1}}
\newcommand{\pv}{{\mathbf p}}
\newcommand{\ts}{\tau_{\text{S}}}

\usepackage{stmaryrd}
\newcommand{\LR}{\overset{{\shortrightarrow}}{L}}
\newcommand{\LL}{\overset{{\shortleftarrow}}{L}}
\newcommand{\SR}{\overset{{\shortrightarrow}}{S}}
\newcommand{\SL}{\overset{{\shortleftarrow}}{S}}

\newcommand{\wshape}[2][]{%
  \tikz[baseline=-0.5ex,scale=0.15,rotate=#1]
    \draw (0:1)
      \foreach \k in {1,...,#2} { -- (\k*360/#2:1) } -- cycle;}
\newcommand{\octagont}[1][]{\wshape[#1]{8}}
\newcommand{\dodecagont}[1][]{\wshape[#1]{12}}
\newcommand{\hexagont}[1][]{\wshape[#1]{6}}
\newcommand{\squarewt}[1][]{\wshape[#1]{4}}
\newcommand{\circwt}{\tikz[baseline=-0.5ex,scale=0.15]\draw (0,0) circle(1);}

\usepackage[colorlinks=true]{hyperref}
\usepackage{cleveref}

\begin{document}

\title{Tiling by Near Coincidence}

\author{Meshy Ochana}
\author{Ron Lifshitz}%
\email{ronlif@tau.ac.il}
\affiliation{School of Physics and Astronomy, Tel Aviv University, Tel Aviv 69978, Israel}

\date{April 6, 2026}

\begin{abstract}
Moiré patterns of twisted and scaled bilayers have recently emerged as a fertile source of quasiperiodic order in two-dimensional materials. Inspired by these systems, we introduce the \emph{near-coincidence method} for generating quasiperiodic tilings of the plane. The method is intuitive---admitting pairs of nearly coincident points from superimposed layers---yet rigorous, as it maps naturally to the well-established cut-and-project formalism. It reproduces classical tilings, including the Ammann--Beenker, the Niizeki--Gähler, and the square and hexagonal Fibonacci tilings, and also reveals new tilings that are unlikely to arise from conventional constructions. The near-coincidence method is algorithmically simple and already realized in a web-based application that generates tilings from specified layer parameters and coincidence conditions. Future extensions include trilayer systems, where preliminary results yield dodecagonal order with square layers, and very small twist angles, where the method may capture the giant moiré patterns of bilayer and trilayer graphene.
\end{abstract}

\maketitle

\section{Statement of the problem and physical motivation}
\label{Sec:Statement}

Quasiperiodic tilings of the plane---like the famous octagonal (8-fold) Ammann--Beenker tiling~\cite{Beenker82,Ammann92}, decagonal (10-fold) Penrose tiling~\cite{Penrose74}, and dodecagonal (12-fold) Stampfli~\cite{Stampfli86} and Niizeki--G\"ahler~\cite{Niizeki87,Gahler88} tilings---have, for decades, served as canonical mathematical models for two-dimensional quasicrystals, and for the high-symmetry surfaces of three-dimensional quasicrystals. The challenge of constructing such quasiperiodic tilings, with prescribed rotational symmetry, lies in the physical requirement that the collection of vertices form a \emph{Delone set}---that is---be both \emph{uniformly discrete} (so that vertices are never arbitrarily close to one another) and \emph{relatively dense} (so that no arbitrarily large voids appear). 

A naive construction, taking the closure under vector addition of an $n$-fold symmetric star of vectors, yields a valid two-dimensional tiling only when $n=3,4,$ or $6$, in which case the outcome is periodic. For all other values of $n>2$ the point set becomes dense in the plane, thereby violating uniform discreteness. Thus, to achieve a valid Delone set without compromising the rotational symmetry, one must employ a systematic selection procedure that decides which of the candidate points are retained and which must be excluded.

Two rigorous and well-established procedures for doing precisely this are the \emph{dual-grid} and the \emph{cut-and-project} methods, both thoroughly presented in standard textbooks~\cite{Senechal95,Baake13}. In the dual-grid method, the acceptance of vertices into the tiling---and of points into the associated Delone set---is governed by the topology of an appropriately defined grid in a dual space. In the cut-and-project method, by contrast, the closure of the initial $n$-fold symmetric star of vectors is embedded in a higher-dimensional space, where it forms a periodic lattice. Points of this lattice are projected back onto \emph{physical space}, but only if their projection, onto the orthogonal complement subspace, or \emph{internal space}, falls within a well-defined \emph{acceptance domain}, thus producing the quasiperiodic tiling.

In the present work we introduce a new and more intuitive procedure for generating quasiperiodic tilings of the plane, inspired by the recent surge of interest in twisted bilayer graphene~\cite{Andrei20}---which is known to exhibit long-range aperiodic order with hexagonal~\cite{Lai25} and dodecagonal~\cite{Ahn18} symmetry---as well as in other multilayered two-dimensional materials. These physical systems continue to attract attention because of their remarkable tunability, which allows one to engineer properties such as superconductivity~\cite{Bistritzer11,Cao18}, controllable magnetism~\cite{Gonzalez17}, and ferroelectricity~\cite{Vizner21}, together with correlated electronic behavior that clearly reflects the underlying quasiperiodic order~\cite{Uri23}. 

Plainly stated, the procedure is based on a bilayer composed of two identical periodic tilings, or point sets, with one layer rotated or scaled relative to the other. Pairs of points---one from each layer---that nearly coincide are merged into single points. These merged points are then accepted as potential vertices of the tiling, while all remaining points are discarded. In this way, the resulting point set satisfies the Delone conditions, while retaining the desired rotational symmetry.

\setlength{\fboxsep}{0pt} 
\begin{figure}[bht]
  \centering
    \subfloat[]{\label{fig:moire a} \fbox{\includegraphics[width=0.3\columnwidth]{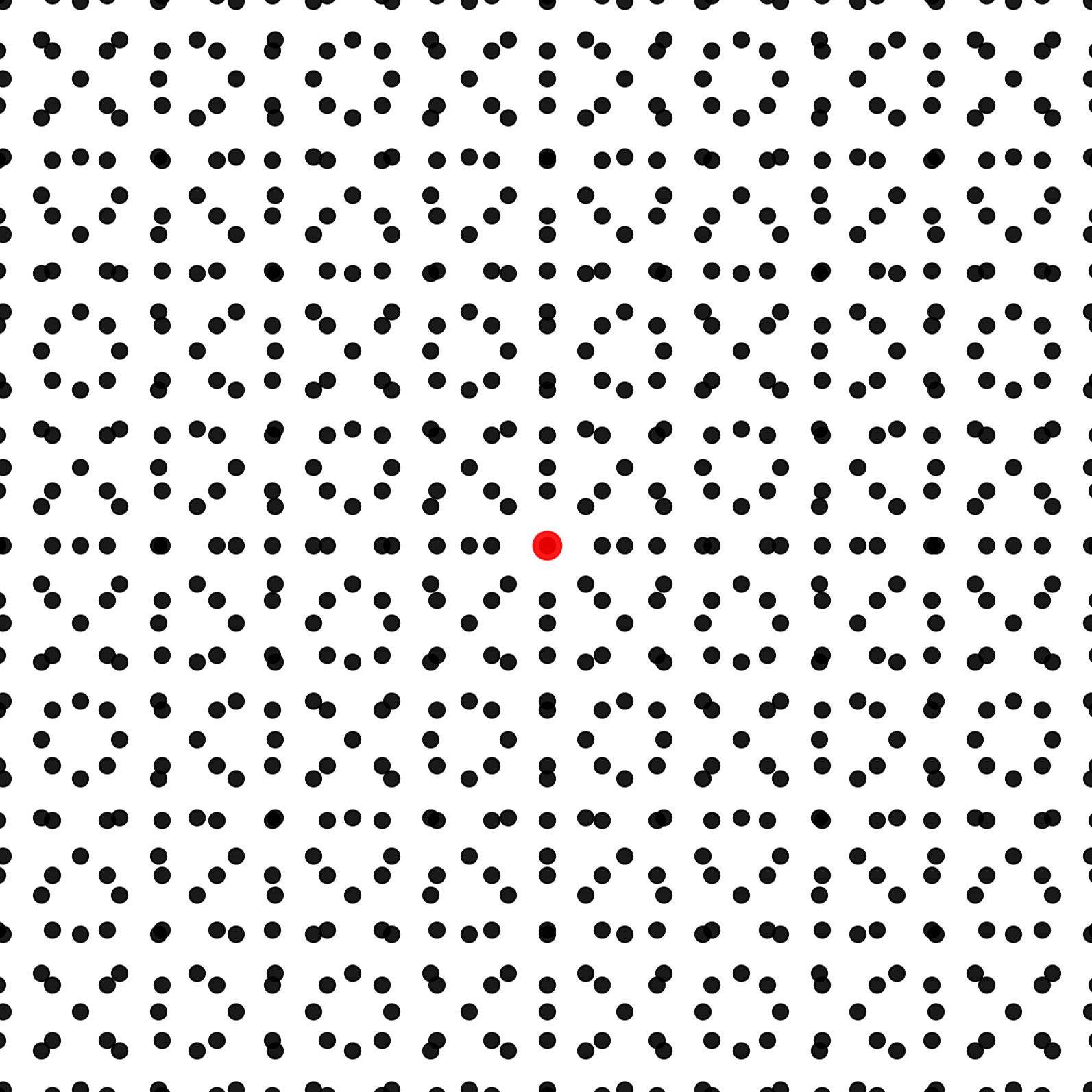}}}
    \hfill
    \subfloat[]{\label{fig:moire b} \fbox{\includegraphics[width=0.3\columnwidth]{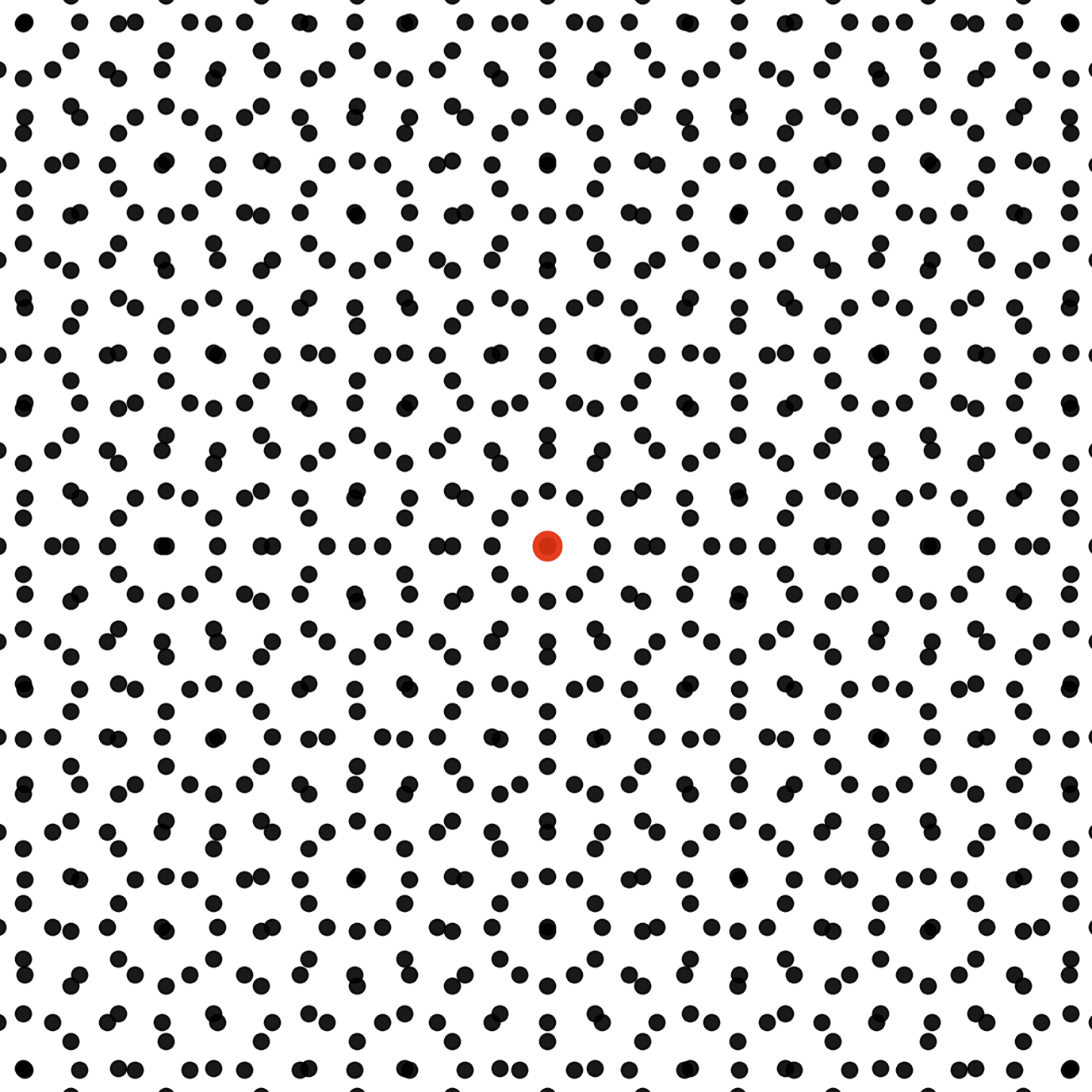}}}
    \hfill
    \subfloat[]{\label{fig:moire c} \fbox{\includegraphics[width=0.3\columnwidth]{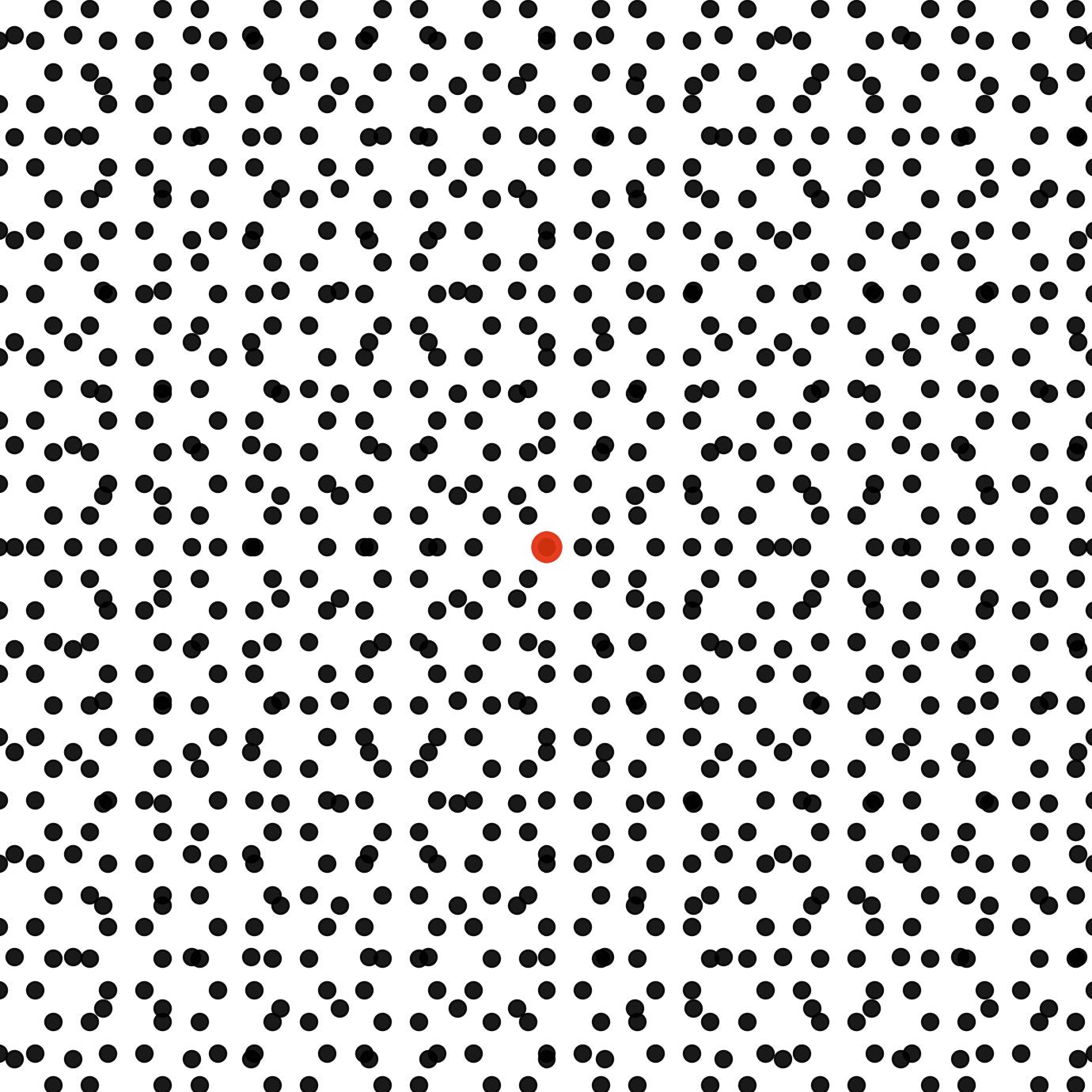}}}
    \caption{Moiré patterns of (a) twisted square-lattice bilayer (rotated by $45^\circ$) producing an octagonal pattern, (b) twisted triangular-lattice bilayer (rotated by $30^\circ$) producing a dodecagonal pattern, and (c) scaled honeycomb-tiling bilayer (scaled by the golden mean  $(1+\sqrt{5})/2$) producing a hexagonal quasiperiodic pattern. A red dot at the center indicates the single point of perfect coincidence of the two layers.}
    \label{fig:moire}
\end{figure}

\begin{figure*}
    \centering
\subfloat[]{\label{fig:steps-a} \includegraphics[width=0.55\textwidth]{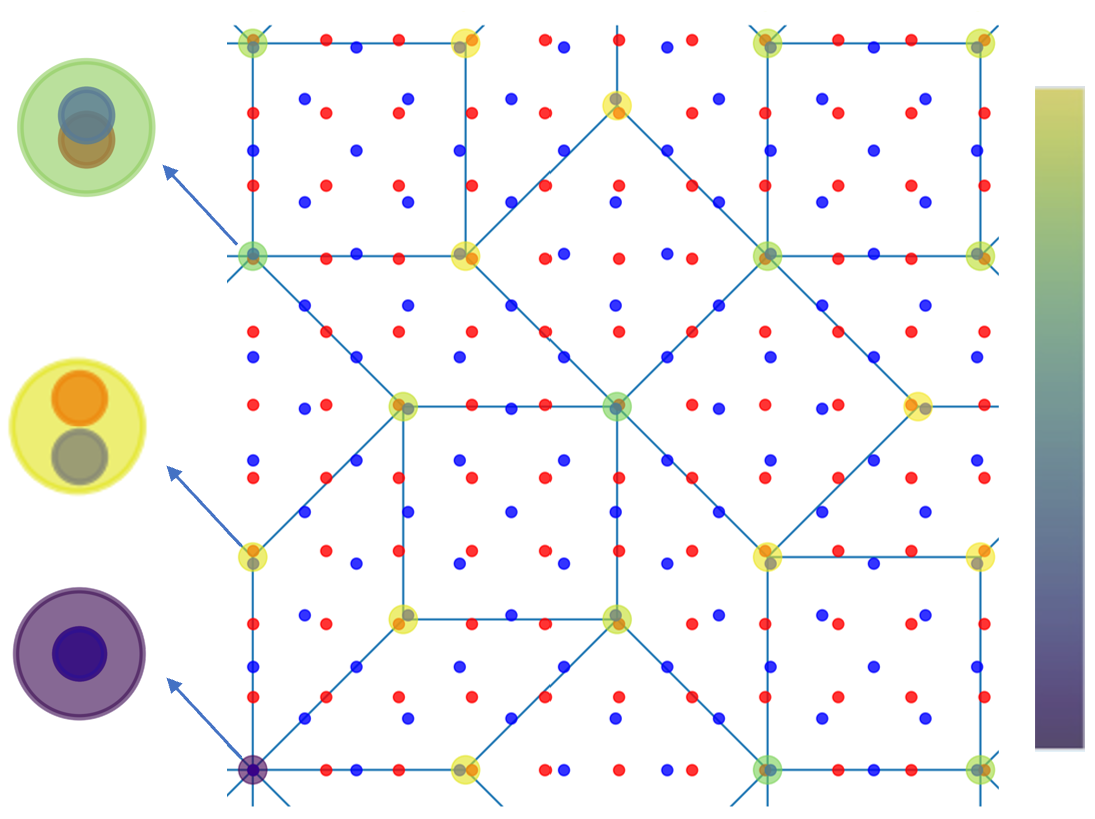}}
\hfill
\subfloat[]{\label{fig:steps-b} \includegraphics[width=0.42\textwidth]{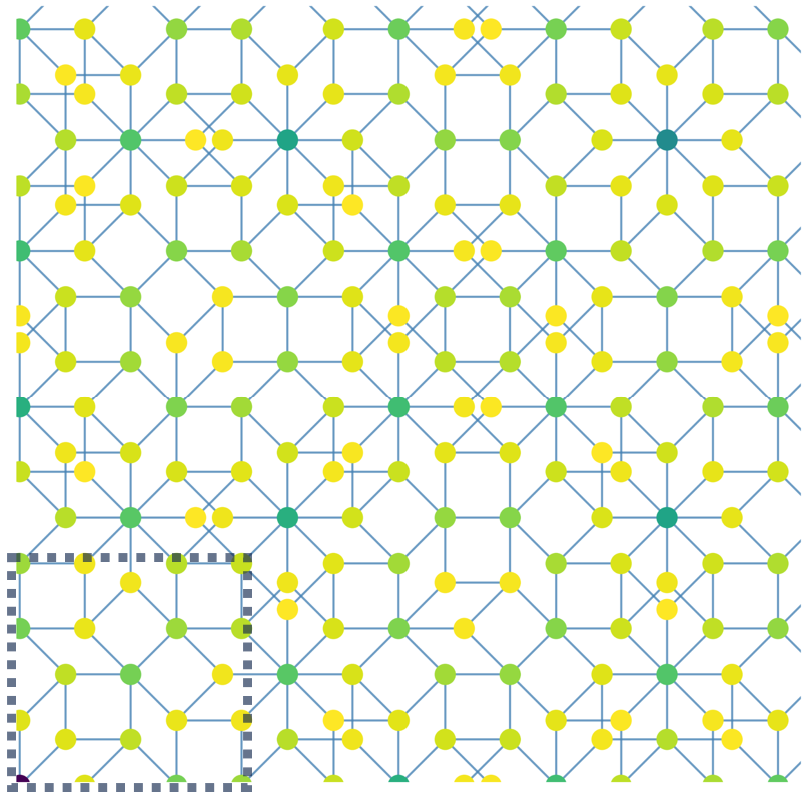}}
\caption{(a) Step-by-step construction of an octagonal tiling by the near-coincidence method: 1.~Two layers of identical periodic square lattices---a blue lattice rotated by $45^\circ$ about the bottom-left point with respect to a red one---are superimposed; 2.~Pairs of points whose separation is shorter than a prescribed threshold are identified as sites of near coincidence, and a potential tiling-vertex is placed at their midpoint, and color coded on a yellow-to-purple scale according to their degree of coincidence; 3.~Edges are drawn between vertices at permitted distances. (b) A larger patch of the constructed tiling, with the section in panel (a) outlined at the bottom-left corner, after removal of the original red and blue points, showing apparent defects consisting of nearby pairs of vertices, and leading to overlapping tiles and crossing edges.}
\label{fig:steps}
\end{figure*}

Figure~\ref{fig:moire} illustrates three typical incommensurate bilayers, which will serve as the paradigmatic examples in what follows. The first is composed of a pair of square lattices, with one lattice rotated by $45^\circ$; the second, of a pair of triangular lattices rotated by $30^\circ$; and the third, of a pair of honeycomb tilings, scaled relative to each other by the golden mean $\tau=(1+\sqrt{5})/2$. Each produces a point set with the corresponding octagonal, dodecagonal, or hexagonal symmetry, yet one that evidently violates uniform discreteness, or the `closeness' condition.

When the rotation angle is commensurate, or alternatively, the scaling factor is rational, the bilayer admits an infinite set of perfect coincidence sites. 
These sites form a periodic sublattice and hence a periodic tiling with an enlarged unit cell, preserving the symmetry of the individual layers up to a reorientation of the point group. A thorough analysis of such commensurate bilayers has recently been provided by Gratias and Quiquandon~\cite{Gratias23}. However, in the generic, incommensurate case---analyzed more recently by the same authors~\cite{Quiquandon25}---at most a single perfect coincidence site can exist. Nevertheless, for any threshold $r>0$, there is always a countable infinity of pairs of nearly coincident points---one from each layer---that are separated by no more than a distance $r$. We argue here that these points of near coincidence provide a natural and intuitive basis for constructing a quasiperiodic tiling with the required symmetry.

To emphasize the intuitiveness of this \emph{near-coincidence} method, we begin in Sec.~\ref{Sec:Generating} with a step-by-step illustration of the construction of an octagonal tiling from two superimposed square lattices rotated by $45^\circ$ (cf.\ Fig.~\ref{fig:moire a}). Informed readers may skip this part and proceed directly to Sec.~\ref{Sec:Ammann}, where we establish the validity of the method by mapping it to the cut-and-project formalism, and by revealing the substitution rules---originally derived by Beenker~\cite{Beenker82}---by varying the coincidence threshold. We then demonstrate the versatility of the approach in Sec.~\ref{Sec:4-dodecagonal}, generating dodecagonal tilings~\cite{Niizeki87,Gahler88} from bilayers of triangular lattices rotated by $30^\circ$ (cf.\ Fig.~\ref{fig:moire b}), and in Sec.~\ref{Sec:Fibonacci}, producing quasiperiodic square~\cite{Lifshitz02} and hexagonal~\cite{Coates24} Fibonacci tilings from bilayers scaled by the golden mean (cf.\ Fig.~\ref{fig:moire c}). We conclude in Sec.~\ref{Sec:Future} with a summary and an outlook toward future directions. 

\section{Generating a Tiling Using the Near-Coincidence Method} 
\label{Sec:Generating}

The near-coincidence construction begins, as shown in Fig.~\ref{fig:moire}, by superimposing two identical periodic tilings, or point sets, and then rotating or scaling one with respect to the other. Figure~\ref{fig:steps-a} provides a zoomed-in view of Fig.~\ref{fig:moire a}, for the case of two square lattices rotated by $45^\circ$, which produces a point set with octagonal symmetry. Two colors---blue and red---are used to distinguish between the two layers, where the bottom-leftmost point shown is the single point of perfect coincidence about which the rotation is performed. The resulting bilayer contains numerous red--blue pairs that lie close to one another. We refer to such pairs as \emph{sites of near coincidence}, and it is precisely these pairs that provide the basis for generating a quasiperiodic tiling, which in this case is expected to be octagonal. 

The guiding principle is straightforward. Pairs of points that lie closer together are more likely to contribute a vertex to the tiling, whereas more widely separated pairs are less likely to do so. The task, therefore, is (A) to define a systematic criterion for deciding which near-coincidence sites are accepted, mapping them consistently into a point set of potential tiling-vertices that satisfy the Delone conditions; (B) to connect these points with appropriately chosen edges; and finally (C) to perform local cleaning, removing excess points or crossed edges according to simple local rules, in order to obtain a proper tiling~\footnote{We note that the latter two steps appear in other tiling constructions, most commonly when one starts with a smooth quasiperiodic field and then identifies a discrete tiling by (A) selecting maxima as potential vertices; (B) connecting them with edges; and (C) resolving local ambiguities. See, for example,~\protect\cite{Barak05,Jagannathan13,*Mace16,Archer22}.}.

\subsection{Choosing a Coincidence Measure} 
\label{sec: 2-measure}

The most natural measure of coincidence is the Euclidean distance between points: the smaller the distance, the stronger the coincidence. Accordingly, we introduce a threshold $r \in \mathbb{R}$ and accept every pair consisting of a point \red{$\mathbf{p}_1$} from the red layer and a point \blue{$\mathbf{p}_2$} from the blue layer whenever $|\red{\mathbf{p}_1}-\blue{\mathbf{p}_2}| < r$. Each such pair is merged into a potential tiling vertex, placed at the midpoint $(\red{\mathbf{p}_1} + \blue{\mathbf{p}_2})/2$, as illustrated by colored circles in Fig.~\ref{fig:steps-a}. The potential tiling-vertices are color coded on a yellow-to-purple scale to reflect the degree of coincidence. The highest-coincidence sites, resulting from pairs of points lying extremely close to one another, are colored purple, while the lowest-coincidence sites, originating from pairs whose separation is closest to the threshold $r$, are colored yellow. Three different color-coded pairs are highlighted in Fig.~\ref{fig:steps-a} to illustrate this range.

This acceptance criterion is equivalent to defining an isotropic \emph{coincidence window}---a circular disk of radius $r$---and retaining all pairs whose separation vector lies within it. Choosing the value of $r$ is to some extent arbitrary: a smaller $r$ yields fewer and sparser accepted vertices, while a larger $r$ yields a higher vertex density. As we shall see later, one may obtain a variety of valid tilings, or scaled versions of the same tiling, simply by varying $r$. 

\subsection{Connecting Vertices with Edges} 
\label{sec: 2-edges}

Once the potential tiling-vertices have been selected, one must decide how to connect them with edges. There are several seemingly arbitrary choices at this stage, which may be influenced by physical or aesthetic considerations. For example, when modeling a physical system within a tight-binding framework, the choice of edge connections may become important, as it can affect both the topology and the relative strengths of hopping processes between nearby sites.

First, the overall scale of edge lengths is imposed by the choice of threshold $r$, which controls the vertex density. Thus, a smaller $r$ yields longer edges, while a larger $r$ yields shorter edges. In addition, special choices of $r$ may allow a minimal number of permitted edge lengths---possibly only one, as in the octagonal example of Fig.~\ref{fig:steps}---while other choices may require additional edge lengths.  

Once the allowed edge lengths are determined, our default, empirically motivated approach is to draw edges between all pairs of vertices separated by the permitted edge lengths. Applied to the octagonal point set with a single allowed edge length, this procedure produces only two prototiles---a square and a $45^\circ$-rhomb---as shown in Fig.~\ref{fig:steps-a}. Of course, one may also obtain a tiling from the selected point set by a \emph{Voronoi diagram}, or by its dual \emph{Delone triangulation}~\cite[sec.~2.2]{Baake13}, but these generic constructions do not in general recover the particular edge set or prototile structure of interest here.

\subsection{Cleaning Up the Tiling} 
\label{sec: 2-remove}

\begin{figure}[tbh]
\centering
\subfloat[]{\label{fig:ambiguity a} \includegraphics[width=0.23\columnwidth]{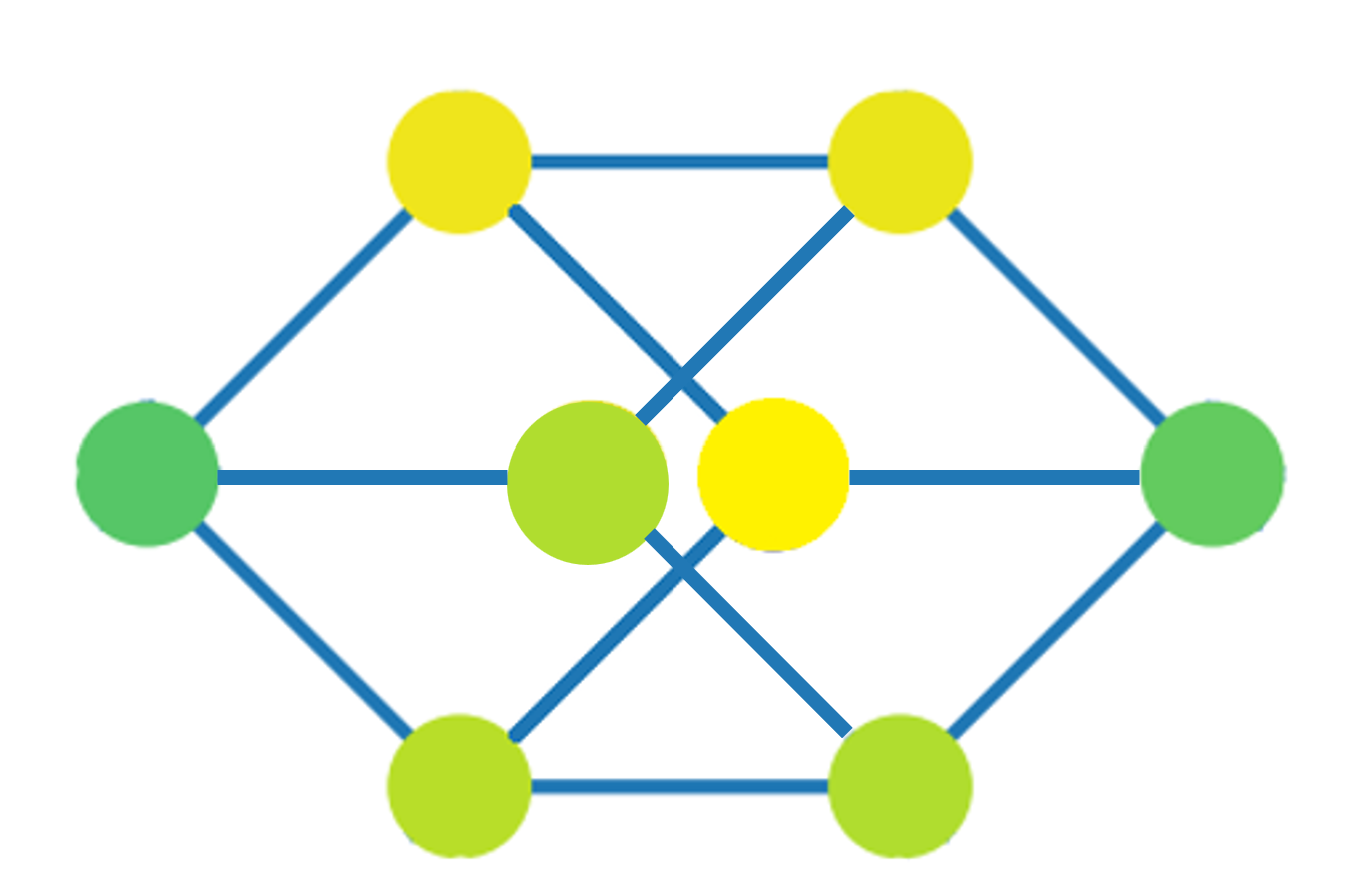}} 
\hfill
\subfloat[]{\label{fig:ambiguity b} \includegraphics[width=0.23\columnwidth]{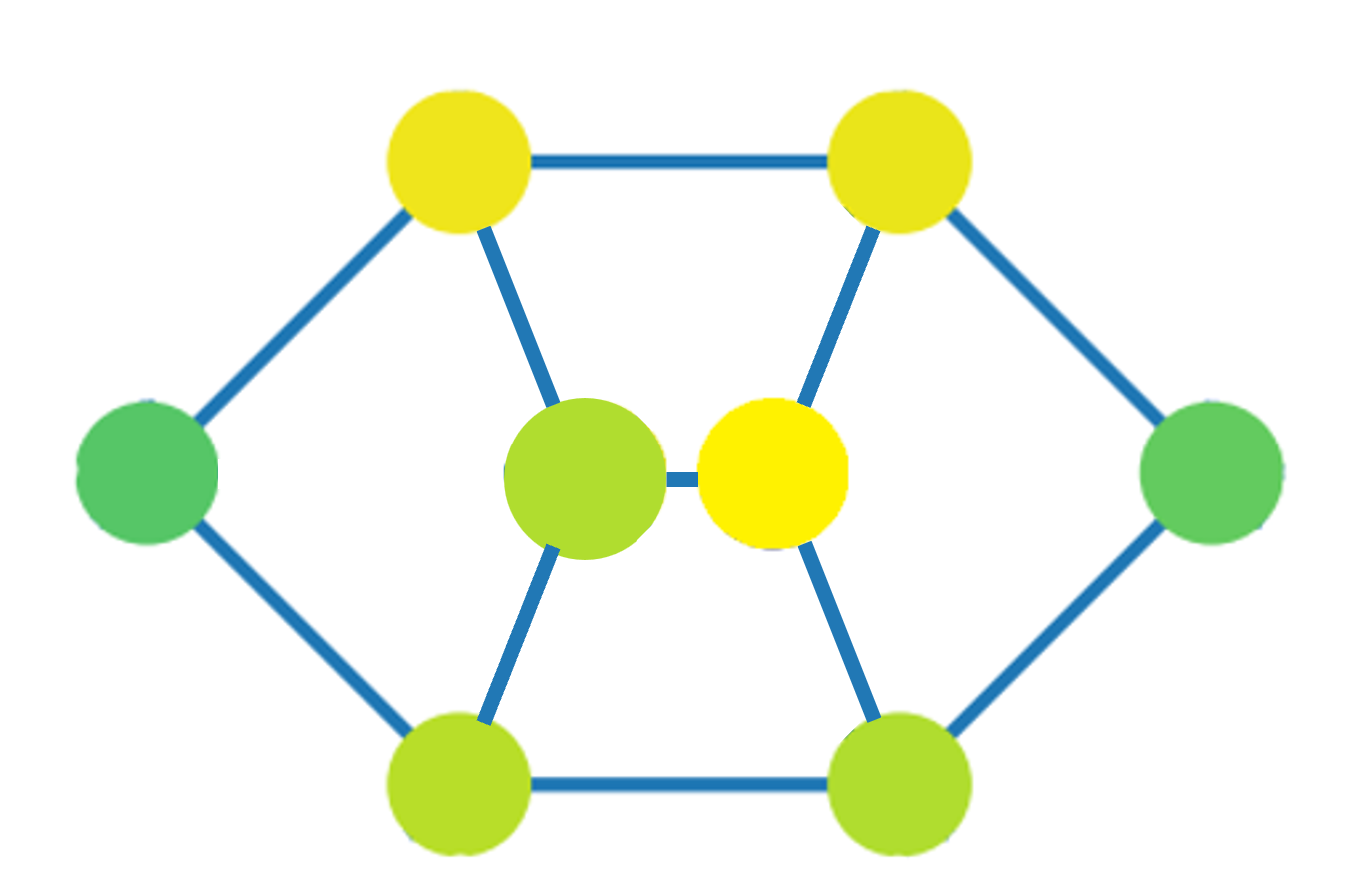}}
\hfill
\subfloat[]{\label{fig:ambiguity c} \includegraphics[width=0.23\columnwidth]{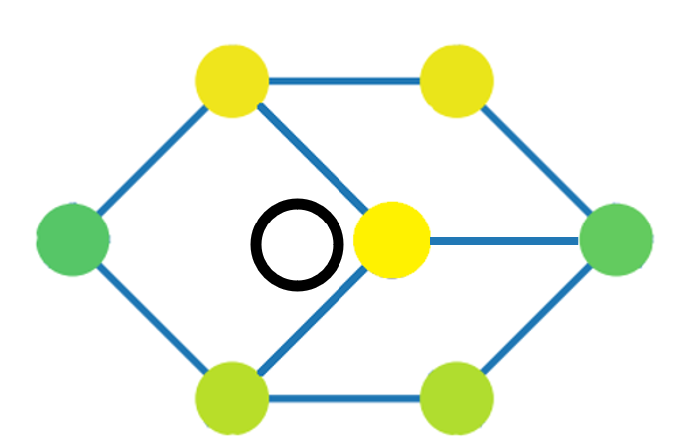}}
\hfill
\subfloat[]{\label{fig:ambiguity d} \includegraphics[width=0.23\columnwidth]{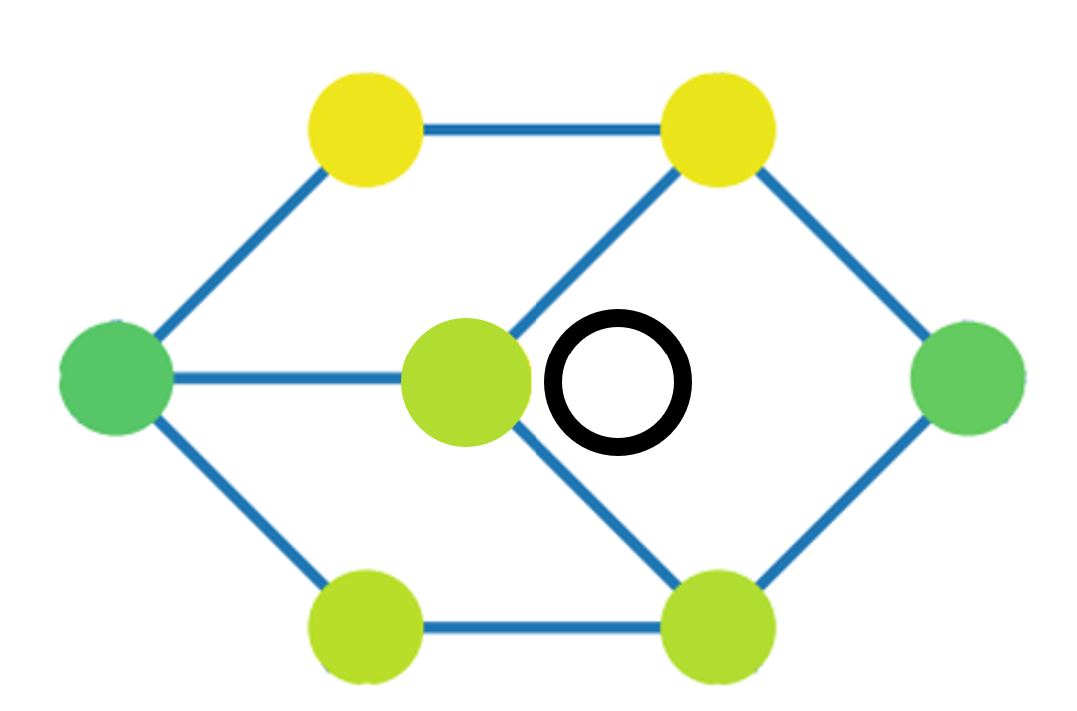}}
\caption{(a) Apparent defect consisting of a pair of nearby vertices, obtained using an isotropic coincidence window. (b) Correcting the defect by introducing two additional edge lengths and two new prototiles---a kite and a trapezoid. (c,d) Two alternative local configurations, consisting of a square and a pair of rhombs, associated with a typical phason flip of the octagonal tiling: in (c) the edge connection is made through the lower-coincidence (yellow) vertex; and in (d) the edge connection is made through the higher-coincidence (green) vertex. One typically chooses the option in (d), discarding the lower-coincidence vertex.}
\label{fig:ambiguity}
\end{figure}

An examination of a larger patch of the tiling, shown in Fig.~\ref{fig:steps-b}, reveals the existence of apparent defects, consisting of pairs of vertices lying too close together and leading to overlapping tiles and crossing edges. A magnified view of one of these defects is shown in Fig.~\ref{fig:ambiguity a}. The yellow-to-green color of the two nearby vertices indicates that they barely make it over the coincidence threshold, and therefore lie on the edge of the coincidence window. There are two typical approaches to correct such defects. The choice between them may be influenced by physical considerations.

\begin{figure*}[bth]
\centering
\subfloat[]{\label{fig:square-kite-trapezoid} \includegraphics[width=0.36\textwidth]{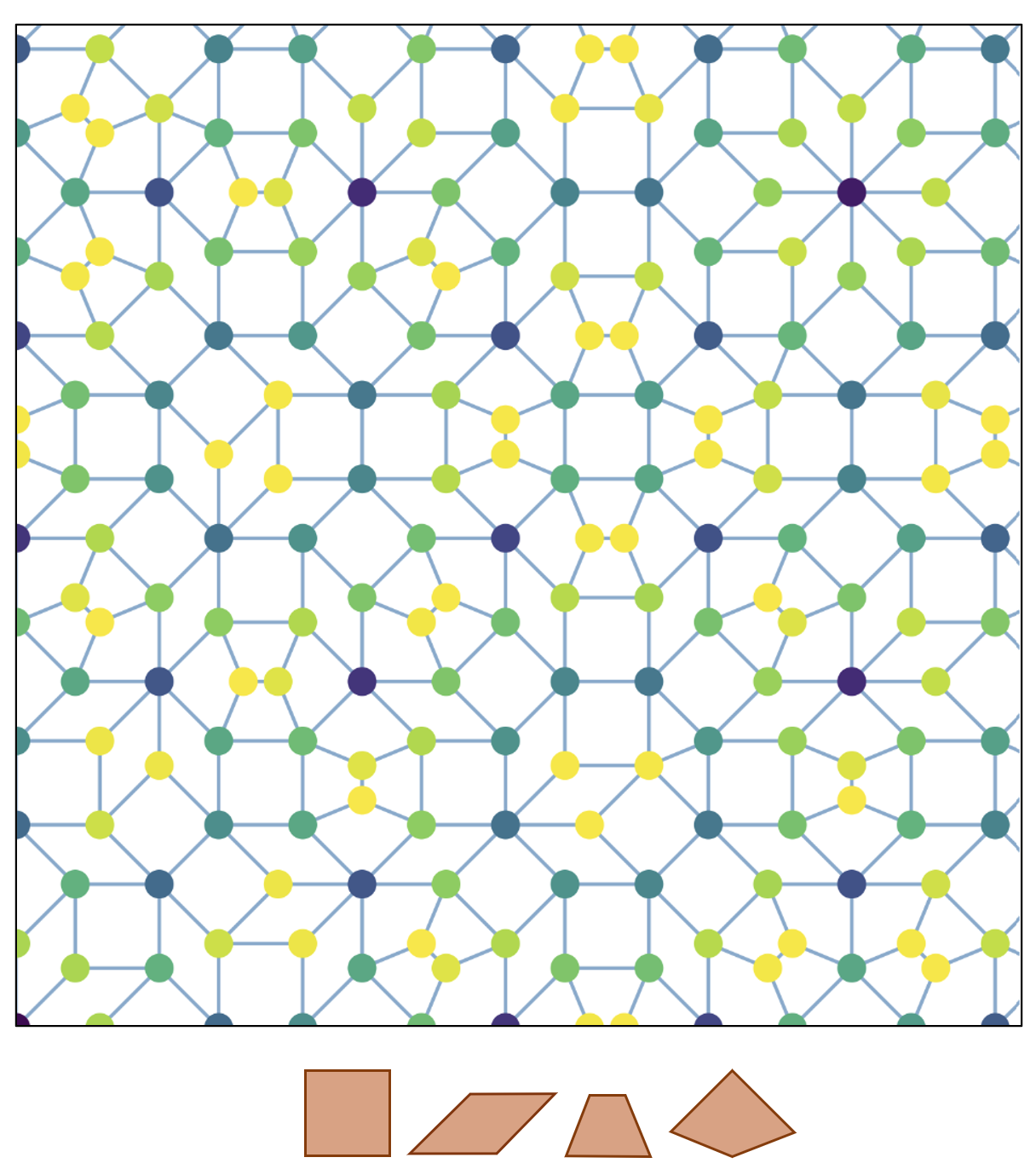}} 
\hfill
\subfloat[]{\label{fig:square-rhomb} \includegraphics[width=0.36\textwidth]{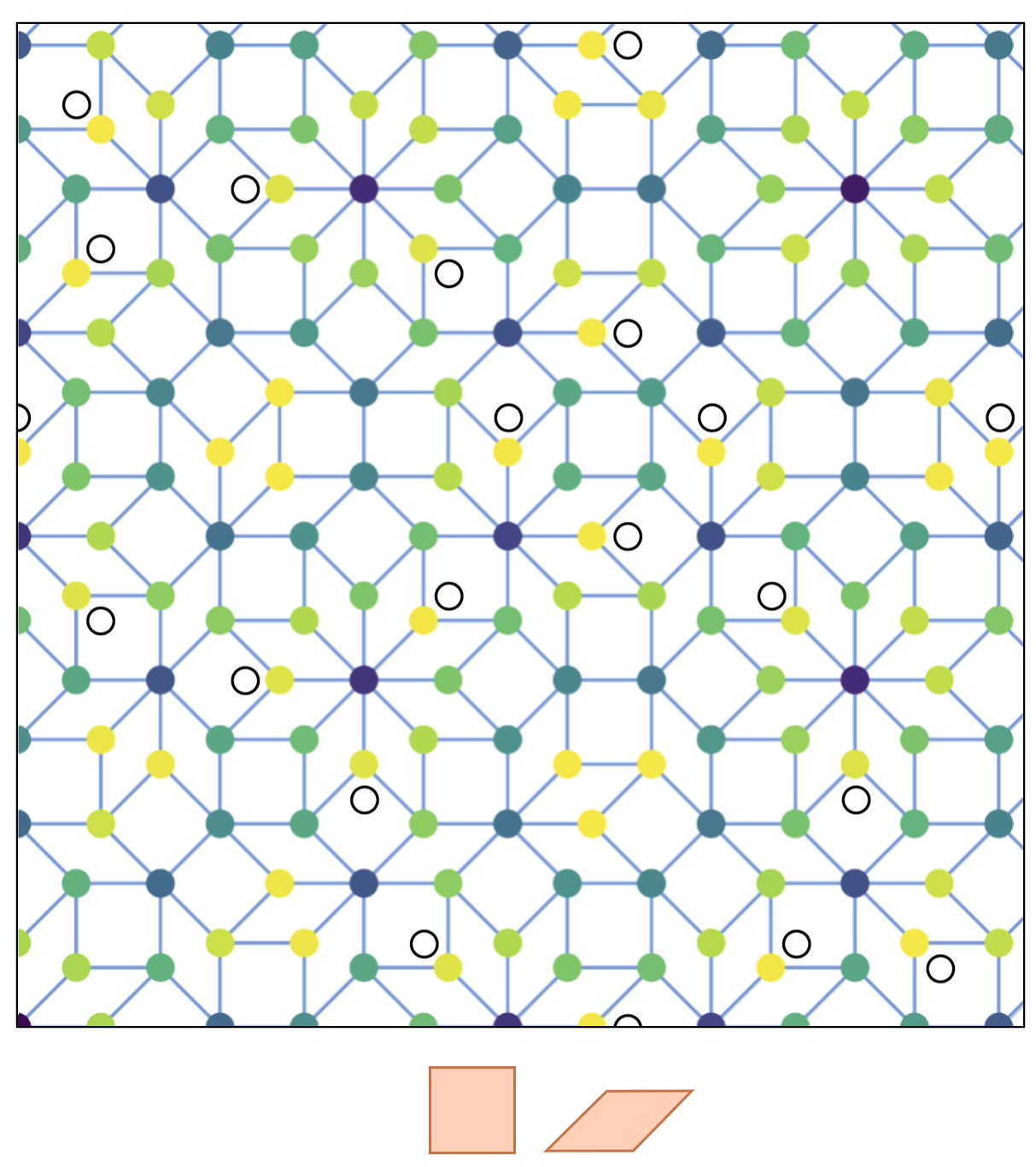}} 
\hfill
\subfloat[]{\label{fig:octagonal-windows}%
  \raisebox{0.1\textwidth}{\includegraphics[width=0.25\textwidth]{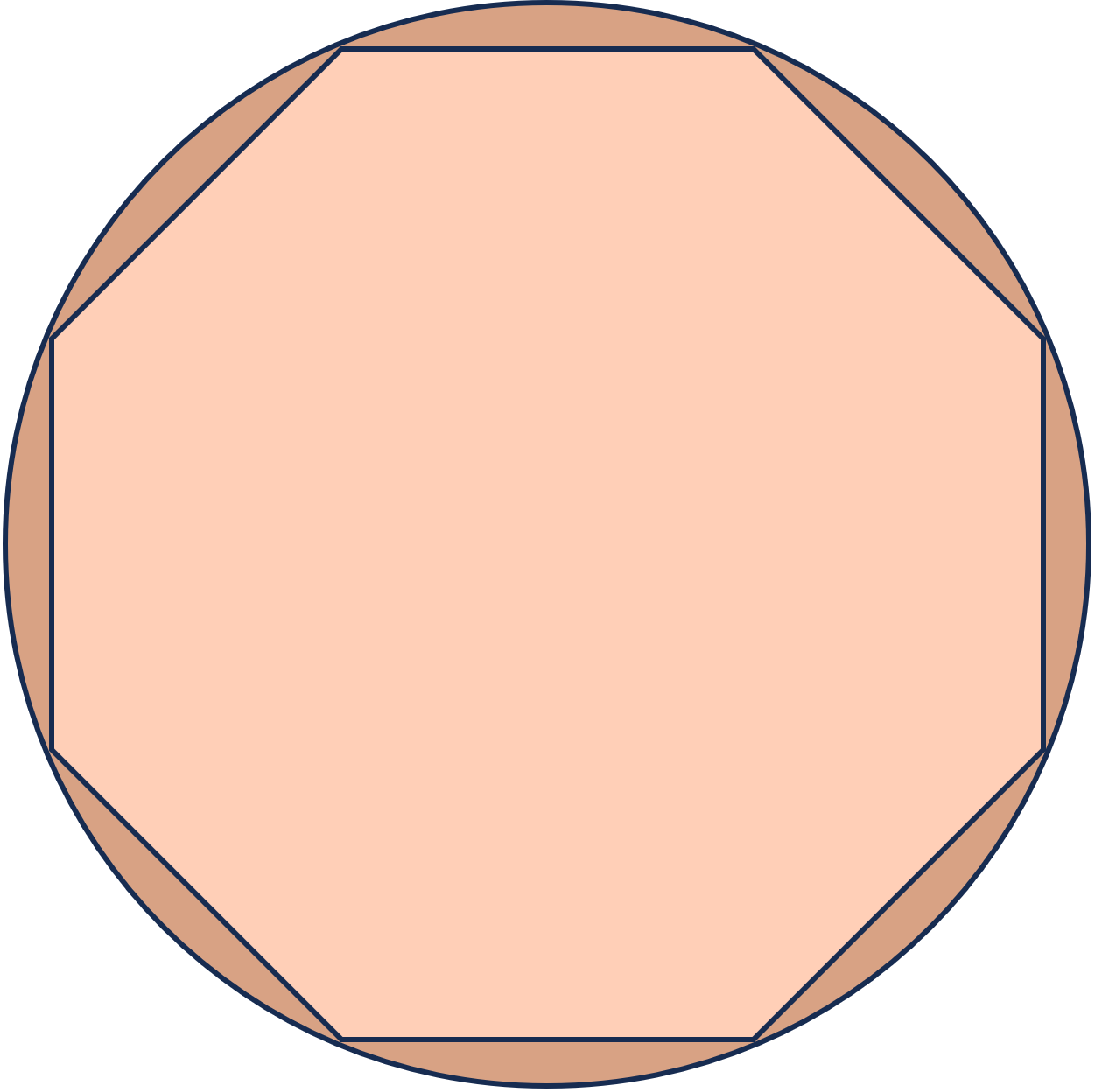}}%
}
\caption{Octagonal tilings obtained by the near-coincidence method as described in Sec.~\ref{Sec:Generating}. (a) The octagonal tiling obtained by admitting all the vertices, shown in Fig.~\ref{fig:steps-b}, whose coincidence falls within the brown circular coincidence window, depicted in panel (c). The tiling requires three different edge lengths, and contains four prototiles---a square, a $45^\circ$-rhomb, a trapezoid, and a kite---displayed beneath the tiling. (b) The octagonal tiling obtained when insisting on a single edge length, thus removing the weaker-coincidence vertex in every pair of nearby vertices. This removes excess points, whose positions are indicated by empty circles, and leads to a tiling containing only the square and the $45^\circ$-rhomb prototiles, displayed beneath the tiling. As demonstrated in Sec.~\ref{Sec:Ammann}, the vertices that are accepted into this tiling correspond to using the light-brown octagonal coincidence window, shown in panel (c), inscribed within the original circular one. This is the well-known Ammann--Beenker~\cite{Beenker82, Ammann92} tiling. (c) Coincidence windows selecting the point sets that are admitted into the octagonal tilings in (a) and (b).}
\label{fig:octagonal-tilings}
\end{figure*}

The first is to admit both of the nearby vertices into the tiling. This requires a relaxation of the single-edge-length preference, allowing two additional edge lengths, and splitting the stretched hexagon that encompasses the pair of nearby vertices into two pairs of additional prototiles---a kite and a trapezoid---as shown in Fig.~\ref{fig:ambiguity b}. This, for example, might be an appropriate choice if there is sufficient space for two atoms to reside so close together. A patch of the final octagonal tiling, after rearranging all the required edges, is shown in Fig.~\ref{fig:square-kite-trapezoid}. 

If there is insufficient space for two atoms at such nearby positions, one must eliminate one of the vertices, splitting the encompassing stretched hexagon into a single square and a pair of rhombs, the two options of which are shown in Figs.~\ref{fig:ambiguity c} and~\ref{fig:ambiguity d}. The natural approach is to prefer the vertex that originates from higher coincidence, corresponding to the closer red--blue pair, and to discard the lower-coincidence vertex. This choice is shown in Fig.~\ref{fig:ambiguity d}. A patch of the final octagonal tiling, after removing all excess points, is shown in Fig.~\ref{fig:square-rhomb}. The informed reader may recognize that this tiling looks very much like the well-known Ammann--Beenker tiling~\cite{Beenker82,Ammann92}. We establish in the following section that this is indeed the case.

It should be noted that redundancies in local tile configurations, associated with small variations in vertex positions like the ones in Figs.~\ref{fig:ambiguity c} and~\ref{fig:ambiguity d}, are not mere mathematical artifacts, but are encountered regularly in physical quasicrystals, where they correspond to so-called \emph{phason flips}~\cite{Lifshitz11}. These are local rearrangements in which atoms may fluctuate between nearly degenerate sites due to thermal excitation, as observed in high-resolution transmission electron microscopy of metallic-alloy quasicrystals~\cite{Edagawa00}. 

\section{Octagonal Tilings by Near Coincidence} 
\label{Sec:Ammann}

\subsection{Mapping to The Cut-and-Project Method} 
\label{sec:3-projection}

The expert reader may immediately notice that the near-coincidence construction admits a natural interpretation within a generalized version of the cut-and-project procedure, mentioned briefly in Sec.~\ref{Sec:Statement}. In this view, one considers the direct product of the two lattices defining the bilayer, and regards pairs of points $(\red{\mathbf{p}_1},\blue{\mathbf{p}_2})$ as elements of a higher-dimensional lattice. The physical-space coordinates are then obtained by a projection that combines the two layers—for example, by taking their midpoint—while the internal-space coordinates are given by their difference $\red{\mathbf{p}_1}-\blue{\mathbf{p}_2}$, whose restriction to a bounded acceptance domain is defined by the coincidence window. In this more general setting, the physical and internal spaces need not be orthogonal, as is often assumed in standard presentations of the cut-and-project method.

At the same time, the formulation adopted here emphasizes the physical origin of the construction, starting from real-space bilayers and their near coincidences, rather than from an abstract higher-dimensional embedding. In what follows, we show explicitly how the near-coincidence method leads directly to the standard cut-and-project description in the octagonal case.

We have constructed the two octagonal tilings of Fig.~\ref{fig:octagonal-tilings} using an isotropic coincidence window, followed by the application of local rules to remove excess vertices and crossed edges. It is natural to examine more closely the set of excess points discarded when insisting on a single edge length. It is evident by direct inspection of Fig.~\ref{fig:square-rhomb} that their positions are well ordered and highly correlated in space. We argued earlier, by noting their yellow-to-green color, that they also do not appear randomly when mapped onto the circular coincidence window. Instead, they accumulate predominantly along the boundaries of the window, exposing their marginal status as low-coincidence sites.

\begin{figure}[tbh]
\centering
\includegraphics[width=0.6\columnwidth]{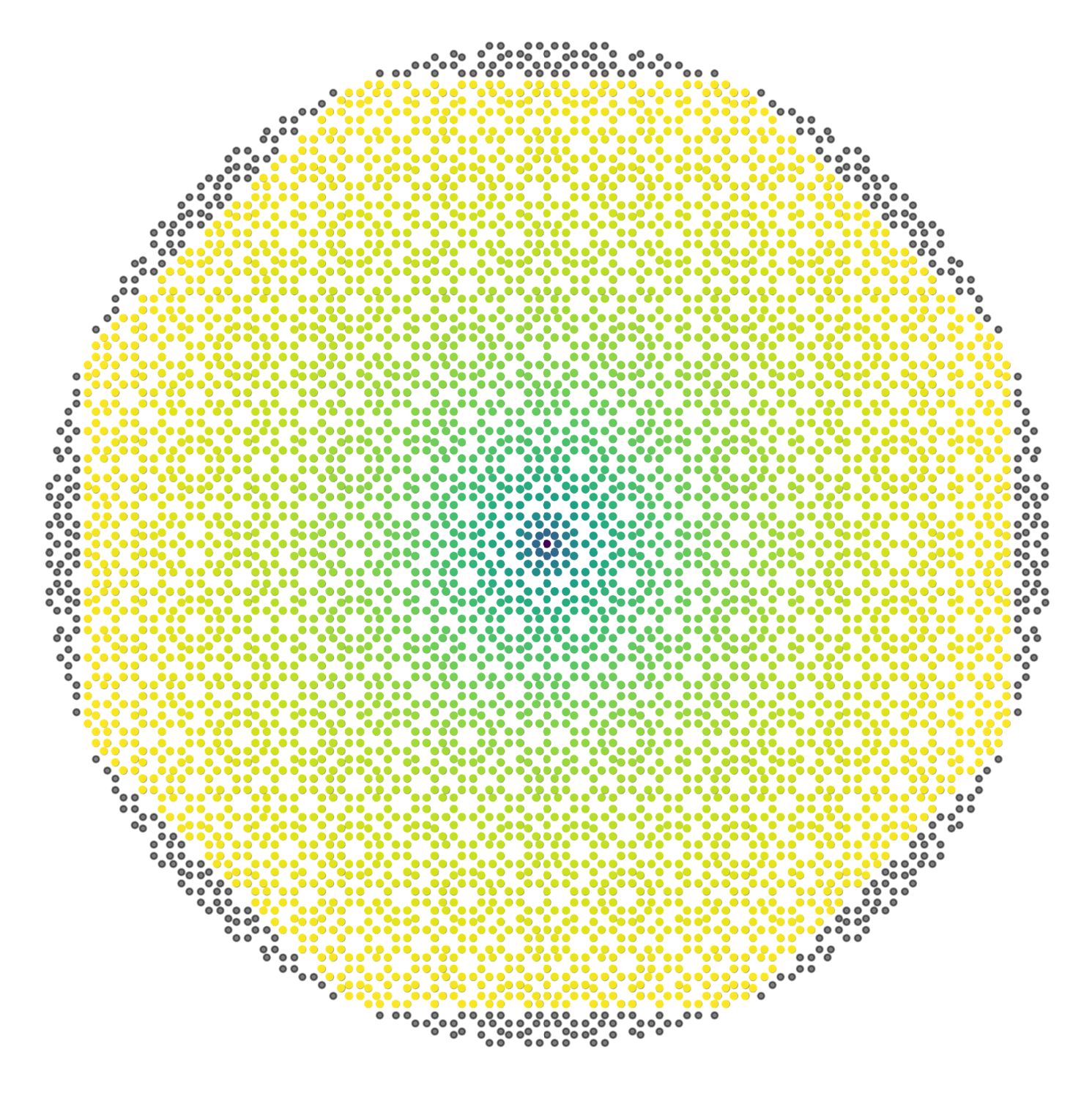}
\caption{Distribution of excess points, mapped onto the centered circular coincidence window, and plotted in black together with the accepted vertices, which are color coded according to their distance from the center. The discarded points outline an octagonal boundary, revealing the octagonal coincidence window.}
\label{fig:excess of points}
\end{figure}

Figure~\ref{fig:excess of points} displays such a mapping of about a thousand tiling vertices. Each vertex is mapped from its position $(\red{\mathbf{p}_1} + \blue{\mathbf{p}_2})/2$ in the tiling to the position of its corresponding separation vector $\red{\mathbf{p}_1} - \blue{\mathbf{p}_2}$, relative to the center of a circular coincidence window. The color coding is maintained for the accepted vertices, while the discarded excess vertices are drawn in black. The distribution of the excluded points outlines, with notable clarity, a perfect \emph{octagonal coincidence window}.

This observation suggests that from the outset, one could have chosen a polygonal coincidence window, instead of a trivial isotropic one, thereby excluding all the excess points in advance, and avoiding the local-rule clean-up step. Thus, what appeared in Sec.~\ref{Sec:Generating} as an \emph{ad hoc} pragmatic act of local cleaning can be replaced by an exact global criterion. Of course, one needs to be careful in choosing a proper polygon that does not break the original symmetry of the bilayer. If one uses a square coincidence window, for example, one obtains a tetragonal, or square, quasiperiodic tiling, instead of the octagonal one.

\begin{figure}[tb]
\centering
\subfloat[]{\label{fig:star oc} \includegraphics[width=0.48\columnwidth]{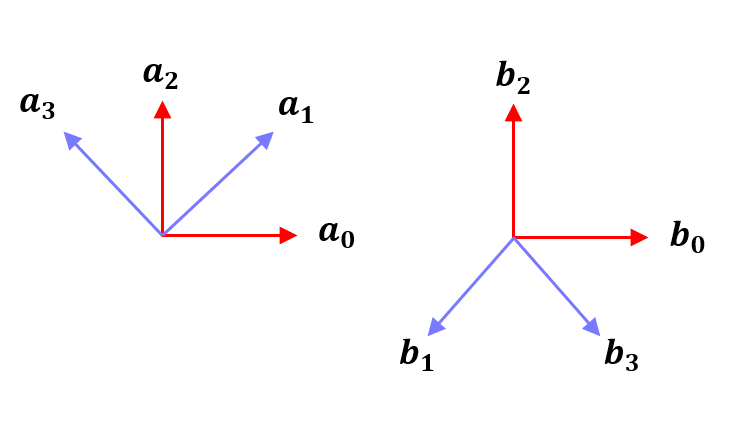}}
\hfill
\subfloat[]{\label{fig:star dod} \includegraphics[width=0.48\columnwidth]{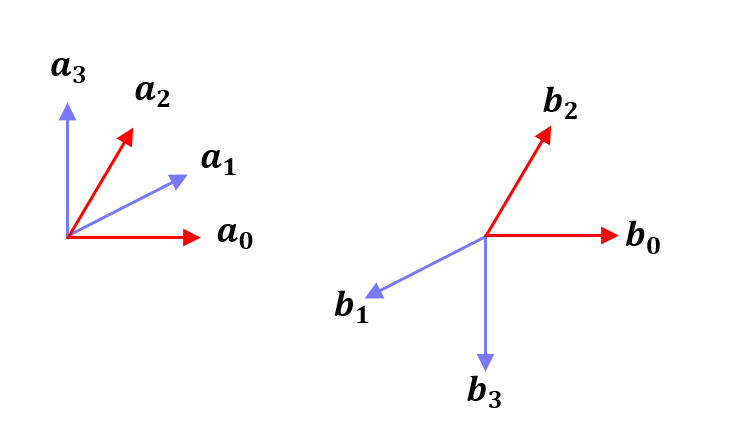}}
\caption{One of the standard star maps used in the cut-and-project method for (a) octagonal; and (b) dodecagonal tilings.}
\label{fig:Star-map}
\end{figure}

At this point, the astute reader may recognize the resemblance of the octagonal coincidence window to the octagonal \emph{acceptance domain}, or so-called \emph{atomic surface}, used in the cut-and-project construction of the Ammann--Beenker tiling. It turns out that this is not accidental, as there is a deep connection between the cut-and-project method and the near-coincidence method. To see this, let us formally describe the set $\TT$ of all vertices in the octagonal tiling by
\begin{equation}\label{Eq:near-coincidence-condition}
\TT = \left\{ \frac{\red{\mathbf{p}_1} + \blue{\mathbf{p}_2}}{2} \;\middle|\; \red{\mathbf{p}_1} - \blue{\mathbf{p}_2} \in \text{CW} \right\},
\end{equation}
where CW stands for the coincidence window, either a circle for the tiling of Fig.~\ref{fig:square-kite-trapezoid}, or an octagon for the tiling of Fig.~\ref{fig:square-rhomb}, both shown together in Fig.~\ref{fig:octagonal-windows}. The points \red{$\mathbf{p}_1$} of the red layer are given by the ${\mathbb Z}$-span of the two red vectors, $\av{0}$ and $\av{2}$, shown in Fig.~\ref{fig:star oc}, while the points \blue{$\mathbf{p}_2$} of the blue layer are given by the ${\mathbb Z}$-span of the two blue vectors, $\av{1}$ and $\av{3}$, where we have defined
\begin{equation}\label{Eq:avectors}
    \av{k} = \left(\cos{k\theta_N},\sin{k\theta_N}\right), 
    \quad \text{with} \quad \theta_N=2\pi/N,
\end{equation}
and here $N=8$.

Accordingly, the tiling vertices can be expressed as
\begin{equation}
\frac{\red{\mathbf{p}_1} + \blue{\mathbf{p}_2}}{2} = \frac{1}{2}\sum_{k=0}^{3} n_k \mathbf{a}_k,
\quad n_k\in{\mathbb Z},
\end{equation}
subject to the condition that the corresponding separation vectors
\begin{equation}
\red{\mathbf{p}_1} - \blue{\mathbf{p}_2} = (n_0\mathbf{a}_0+n_2\mathbf{a}_2) - (n_1\mathbf{a}_1+n_3\mathbf{a}_3) \equiv \sum_{k=0}^{3} n_k \mathbf{b}_k
\end{equation}
are in the coincidence window. Here we have defined a set of four additional two-dimensional vectors, also shown in Fig.~\ref{fig:star oc}, as
\begin{equation}\label{Eq:oct-starmap}
\mathbf{b}_0=\mathbf{a}_0, \quad \mathbf{b}_1=-\mathbf{a}_1, \quad \mathbf{b}_2=\mathbf{a}_2, \quad \mathbf{b}_3=-\mathbf{a}_3.
\end{equation}

At first sight this may seem like a convenience. Yet, up to a factor of 2, it coincides exactly with one of the standard so-called \emph{star maps}, used in the cut-and-project formalism to decide whether a particular integer linear combination of an 8-fold symmetric star of vectors belongs in the tiling (see, for example, Fig.~3a of Socolar~\cite{Socolar89}). The star map associates with every potential vertex $\pv$, of the Ammann--Beenker tiling, a unique two-dimensional orthogonal complement, $\pv^\star$, given by
\begin{equation}\label{Eq:StarMap}
    \star:\quad \pv=\sum_{k=0}^3 n_k \av{k}
    \quad \longmapsto \quad
    \pv^\star=\sum_{k=0}^3 n_k \bv{k}.
\end{equation}
The vertex $\pv$ is accepted into the tiling if its star map $\pv^\star$ belongs to the acceptance domain AD. Thus, the tiling defined in Eq.~\eqref{Eq:near-coincidence-condition} is equivalent, to within a factor of 2, to the tiling defined using the star map,
\begin{equation}
\mathcal{T} = \left\{ \tfrac{1}{2}\sum_{k=0}^{3} n_k \mathbf{a}_k \;\middle|\; \sum_{k=0}^{3} n_k \mathbf{b}_k \in \text{AD} \right\}.
\end{equation}

The star map of the cut-and-project method finds a natural reinterpretation in the near-coincidence framework as a direct distance measure between paired points in the bilayer. The \emph{coincidence window} is therefore a real-space manifestation of the \emph{acceptance domain}. This equivalence---first hinted at by the outline of the discarded points---shows that the near-coincidence method reproduces, in a direct and transparent way, the well-known Ammann--Beenker construction. 

As an additional observation, we have discovered that replacing the octagonal acceptance domain of the Ammann--Beenker tiling by its inscribing circle yields a new octagonal tiling with square, $45^\circ$-rhomb, kite, and trapezoid prototiles. We should note that circular or nonpolygonal windows have been considered in extensions of the cut-and-project method~\cite{Masakova03}; here, they arise naturally as the simplest choice within the bilayer perspective.

\begin{figure}[tb]
\centering
\subfloat[]{\label{fig:substitution a} \includegraphics[width=0.55\columnwidth]{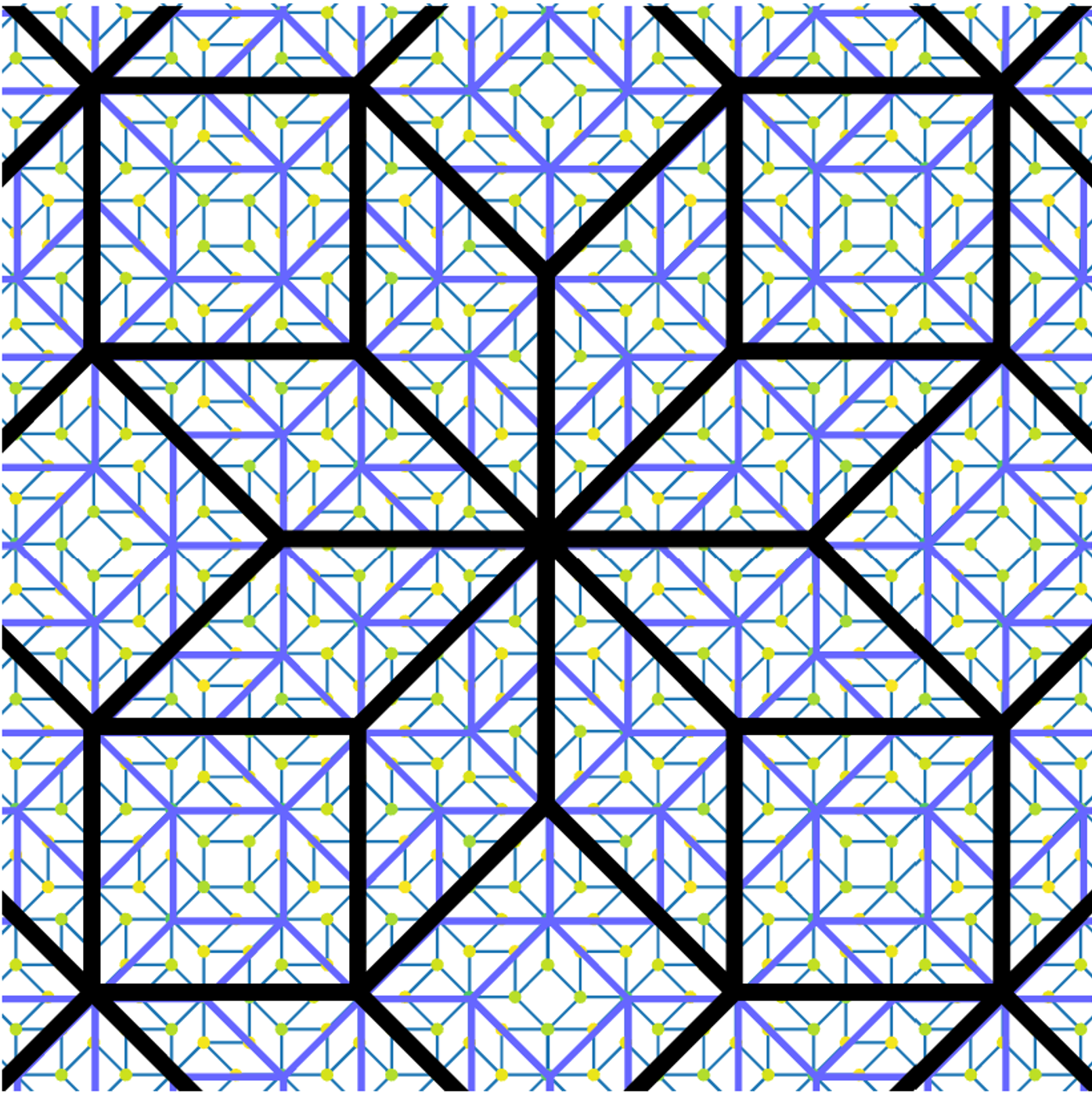}}
\hfill
\subfloat[]{\label{fig:substitution b} \includegraphics[width=0.40\columnwidth]{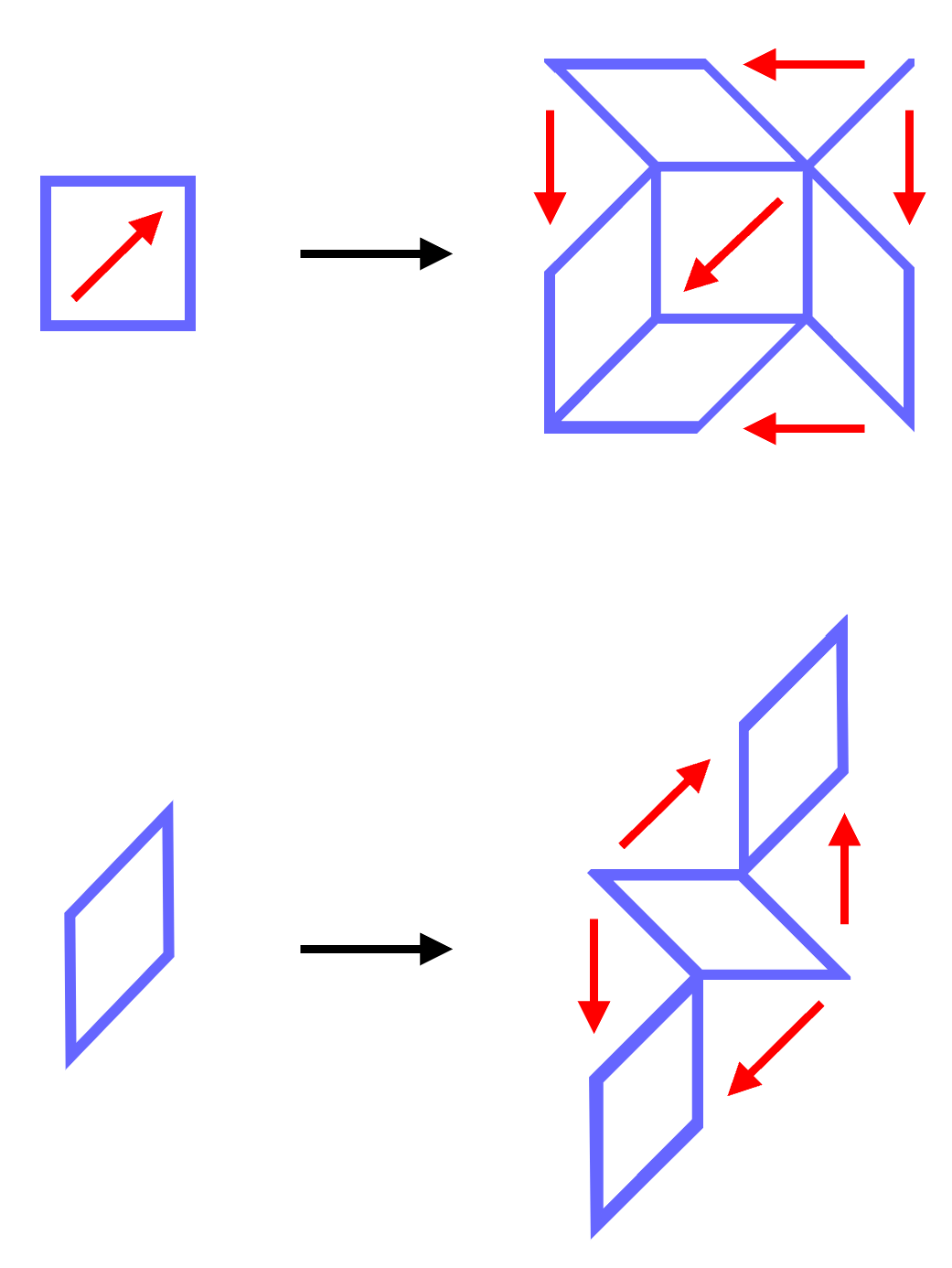}}
\caption{Scaling the octagonal coincidence window twice by the silver-mean $\ts=1+\sqrt{2}$ generates (a) a sequence of three self-similar Ammann--Beenker tilings; which reveals (b) the Ammann--Beenker substitution rules. Note that the square tile has an orientation, as indicated by the red arrows, which has to be maintained throughout the substitution process.}
\label{fig:Substitution rules}
\end{figure}

\subsection{Substitution Rules} 
\label{sec:3-substitution}

As stated earlier, adjusting the size of the coincidence window---equivalently, tuning the threshold radius $r$---controls the vertex density of the tiling. A shorter radius (higher threshold) yields fewer vertices, while a longer radius (lower threshold) yields more. A particularly significant case arises when the threshold is increased, shrinking the linear dimensions of the coincidence window by powers of the so-called inflation factor of the tiling, producing self-similar copies of the tiling. A direct inspection of the Ammann--Beenker tiling reveals that its inflation factor is given by the silver-mean $\ts=1+\sqrt{2}$. This is demonstrated in Fig.~\ref{fig:substitution a} showing the original tiling in thin blue edges, a second tiling in thicker purple edges scaled by a factor of $\ts$, and a third in thick black edges, scaled by a factor of $\ts^2$.

In this way, the known substitution rules of the Ammann--Beenker tiling~\cite[Figs.~6.4 and~6.5]{Beenker82} emerge naturally from the near-coincidence framework. Figure~\ref{fig:substitution b} shows them schematically: the rhomb subdivides symmetrically into three smaller rhombs and four half-squares, whereas the square subdivides asymmetrically into a central square surrounded by four rhombs and four half-squares. The alternating orientations of the squares, indicated by red arrows, can be read directly from the geometry of successive inflated generations in Fig.~\ref{fig:substitution a}. This provides yet another confirmation that the near-coincidence method faithfully reproduces the well-known standard constructions.

\section{Dodecagonal Tilings by Near Coincidence} 
\label{Sec:4-dodecagonal}

The near-coincidence method provides a versatile framework for varying the size, shape, and orientation of the coincidence window, thereby producing a wide variety of tilings. By exploiting its correspondence with the cut-and-project method, one can reproduce familiar tilings using polygonal coincidence windows that mirror the polygonal acceptance domains of the projection approach. Indeed, dodecagonal tilings generated from a bilayer of two triangular lattices twisted by $30^\circ$---shown earlier in Fig.~\ref{fig:moire b}---are related to their cut-and-project counterparts exactly as in the octagonal case. The star-map of Eq.~\eqref{Eq:StarMap} is defined exactly as in Eq.~\eqref{Eq:oct-starmap}, except that now $N=12$, so the $\av{i}$ vectors of Eq.~\eqref{Eq:avectors} span a pair of triangular rather than square lattices, as illustrated in Fig.~\ref{fig:star dod} (see, for example, Fig.~2 of Oxborrow and Henley~\cite{Oxborrow93}).

\begin{figure*}[bht]
\centering
\subfloat[]{\label{fig:dod polygon windows} 
\raisebox{0.035\textwidth}{\includegraphics[width=0.22\textwidth]{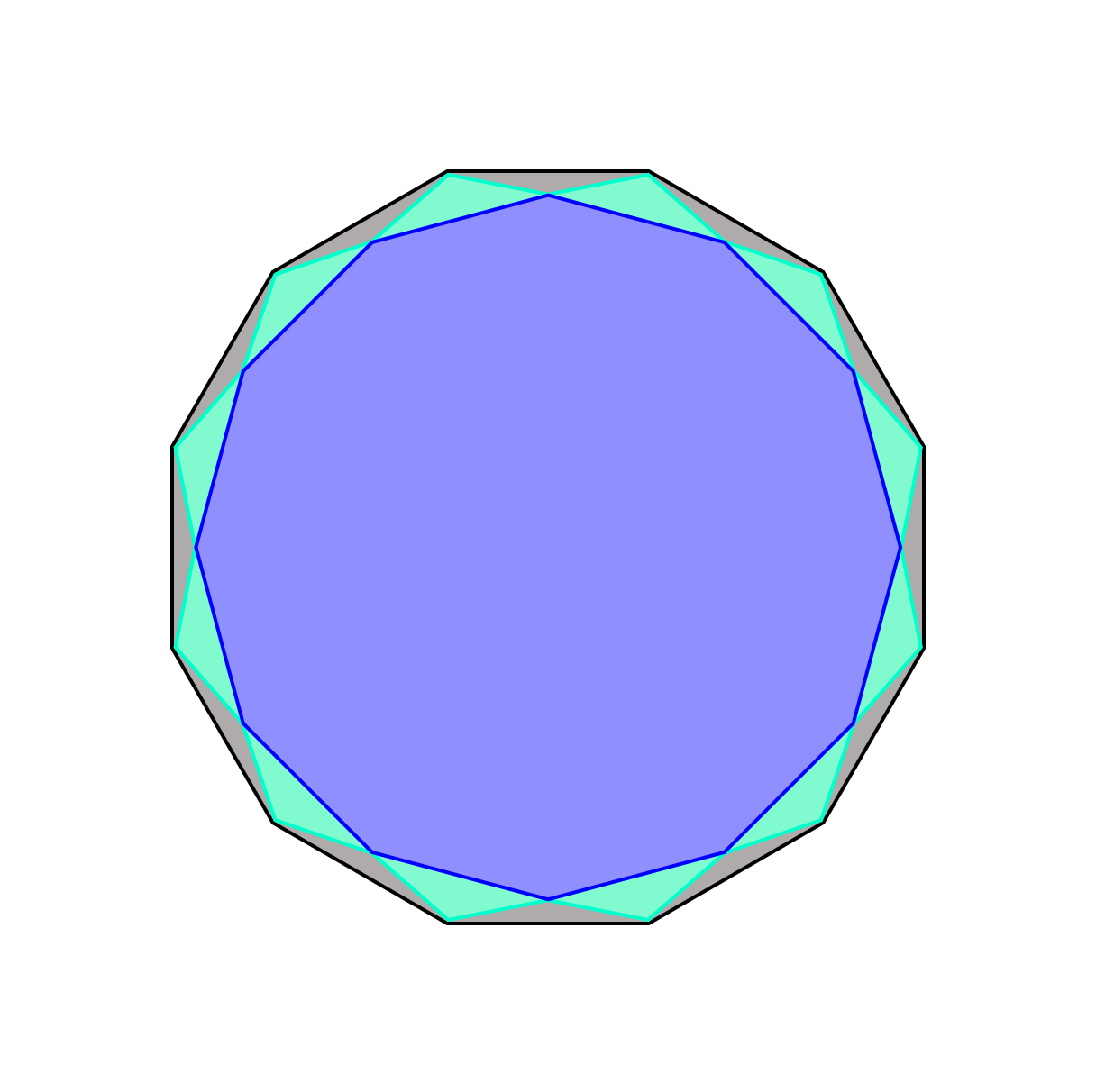}}}
\hfill
\subfloat[]{\label{fig:dod a} \includegraphics[width=0.24\textwidth]{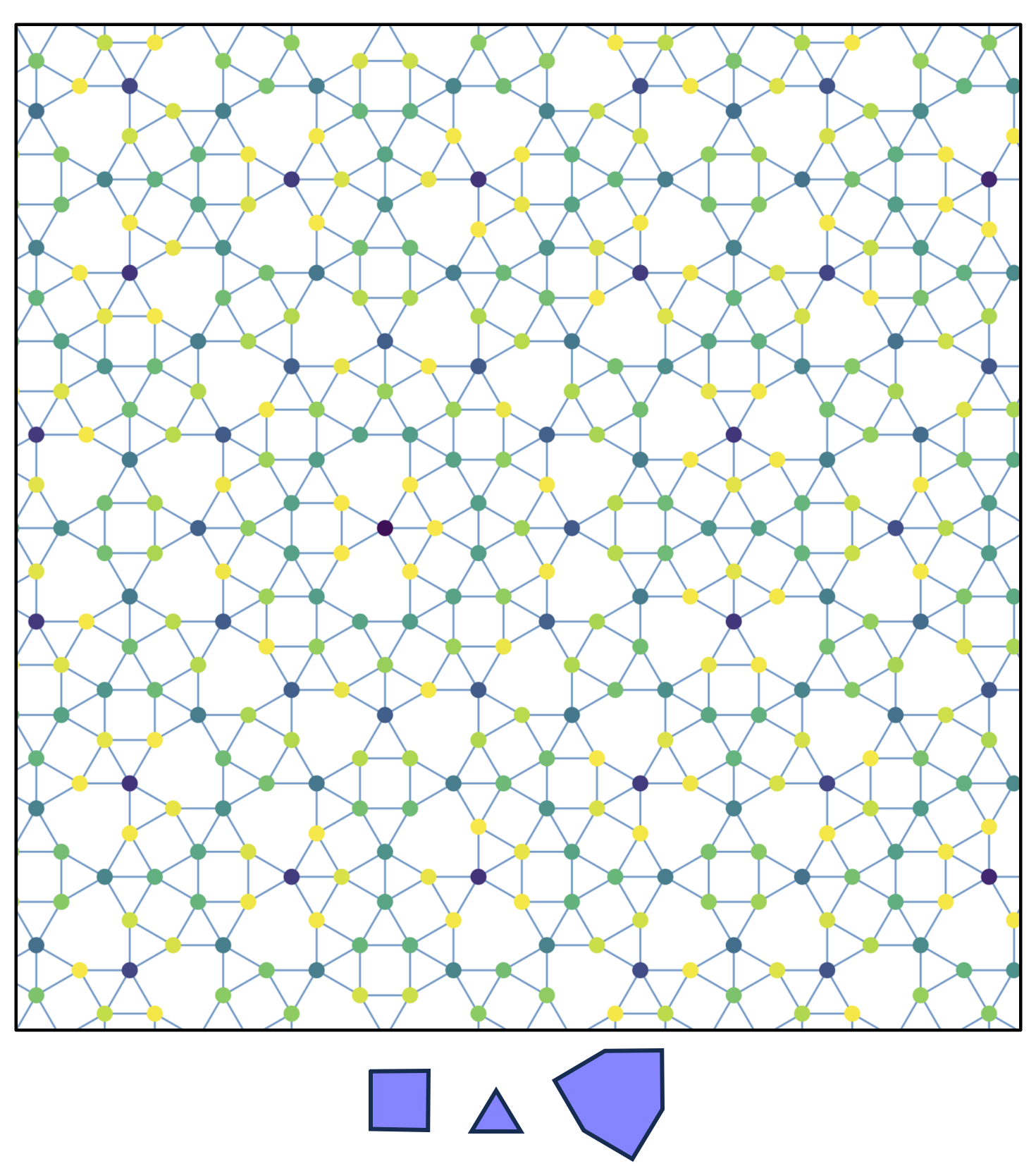}}
\hfill
\subfloat[]{\label{fig:dod b} \includegraphics[width=0.24\textwidth]{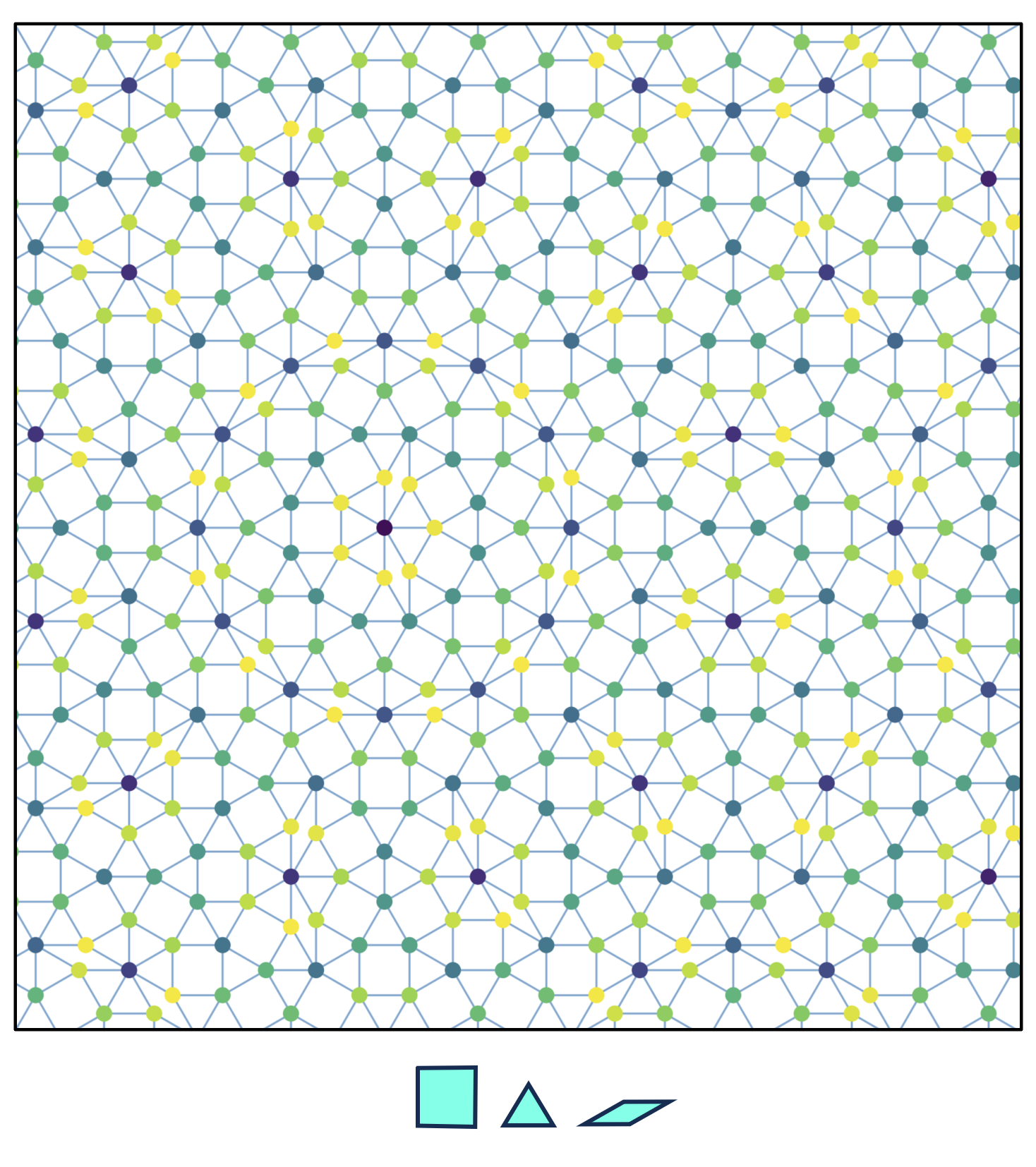}}
\hfill
\subfloat[]{\label{fig:dod c} \includegraphics[width=0.24\textwidth]{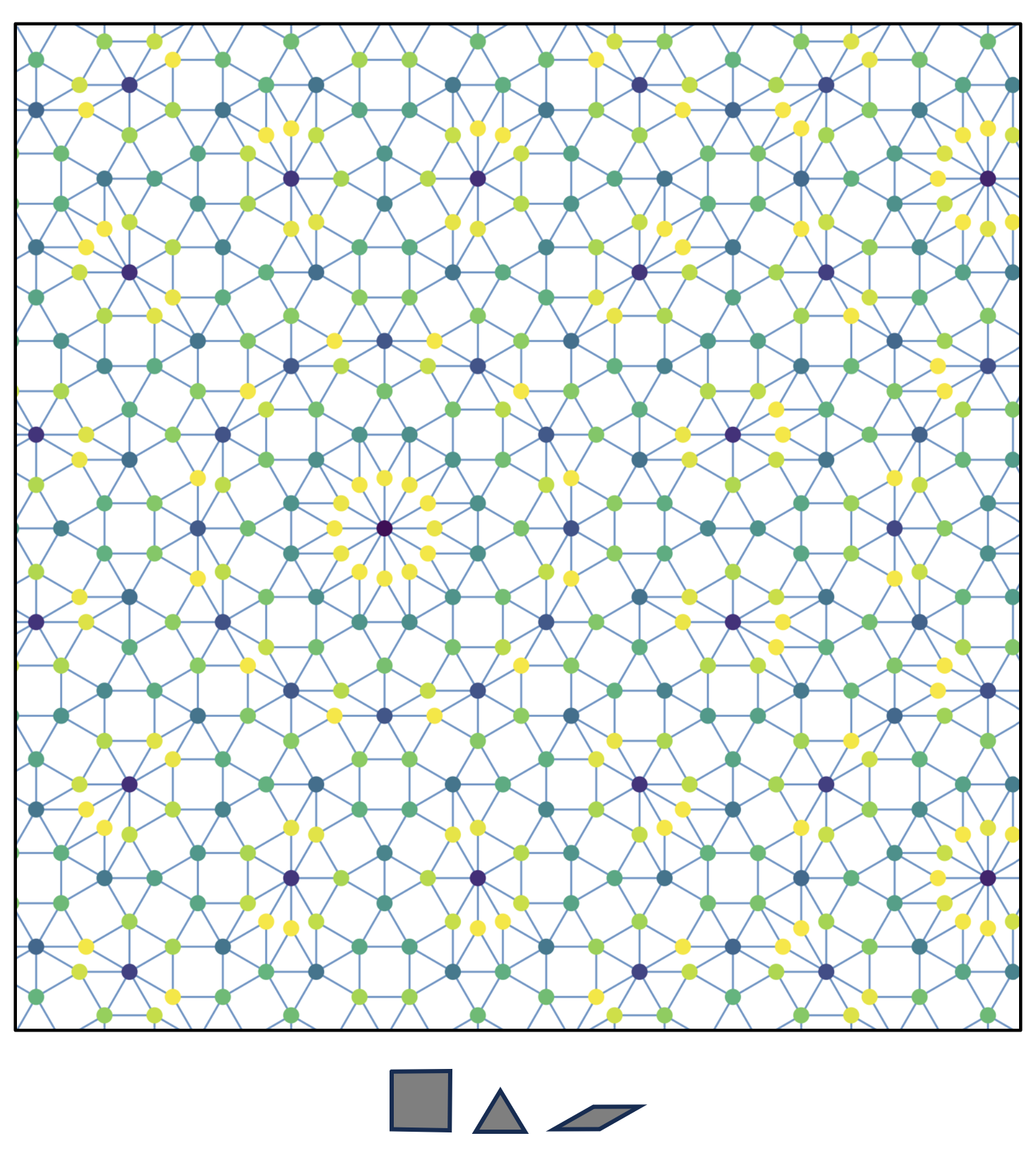}}
\caption{The three dodecagonal Niizeki--Gähler tilings~\cite{Niizeki87,Gahler88}, reproduced by the near coincidence of a $30^\circ$ twisted triangular bilayer, as presented in Fig.~\ref{fig:moire b}. 
(a) Three polygonal coincidence windows with dodecagonal symmetry, used for constructing the displayed tilings. 
(b) The \emph{shield tiling}, generated with the small blue dodecagon, containing three prototiles: the square, the triangle, and the shield. 
(c) The \emph{Niizeki--Gähler tiling}, generated with the nonconvex turquoise 12-fold star, where the shield prototile is replaced by a $30^\circ$ rhomb. 
(d) The \emph{Stampfli tiling}~\cite{Stampfli86}, generated with the large dark-gray dodecagon, retaining the same three prototiles as the Niizeki--Gähler tiling.}
\label{fig:Gahler dodecagonal tilings}
\end{figure*}

With this prescription we reproduce the three Niizeki--Gähler dodecagonal tilings~\cite{Niizeki87,Gahler88} by applying the near-coincidence method to a $30^\circ$ twisted triangular bilayer. The three distinct polygonal coincidence windows used for the construction---equivalent to the polygonal acceptance domains used by Niizeki and Mitani~\cite{Niizeki87} and by Gähler~\cite{Gahler88}---are shown one on top of the other in Fig.~\ref{fig:dod polygon windows}. The smallest window, admitting the lowest density of vertices, is the small blue dodecagon. A slightly larger window is obtained by extending the blue dodecagon into the nonconvex turquoise 12-fold star, formed by the union of a hexagon and its image under a $30^\circ$ rotation. The largest window, admitting the highest density of vertices, is obtained by taking the convex closure of the previous star yielding the large dark-gray dodecagon. 

\begin{figure}[ptbh]
\centering
\subfloat[]{\label{fig:local changes a} 
    \includegraphics[width=0.3\columnwidth]{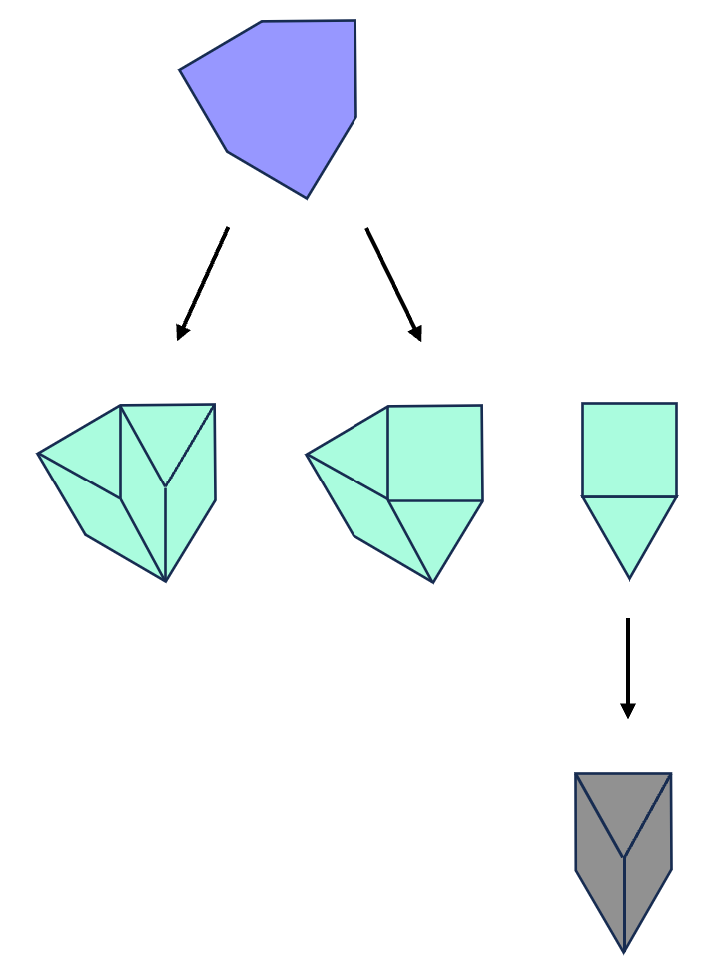}}
\hfill
\subfloat[]{\label{fig:local changes b} 
    \raisebox{5ex}{\includegraphics[width=0.2\columnwidth]{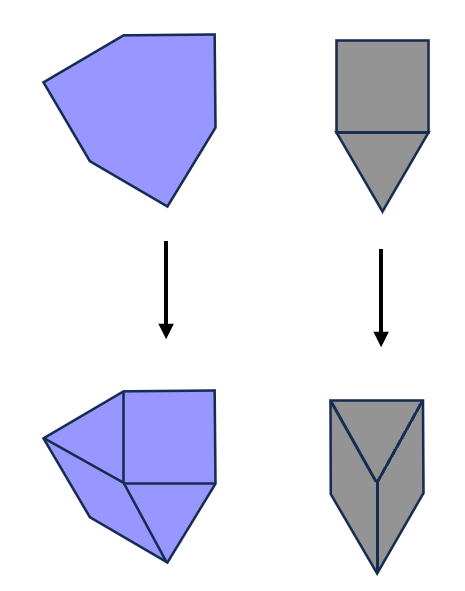}}}
\hfill
\subfloat[]{\label{fig:local changes c} 
    \includegraphics[width=0.29\columnwidth]{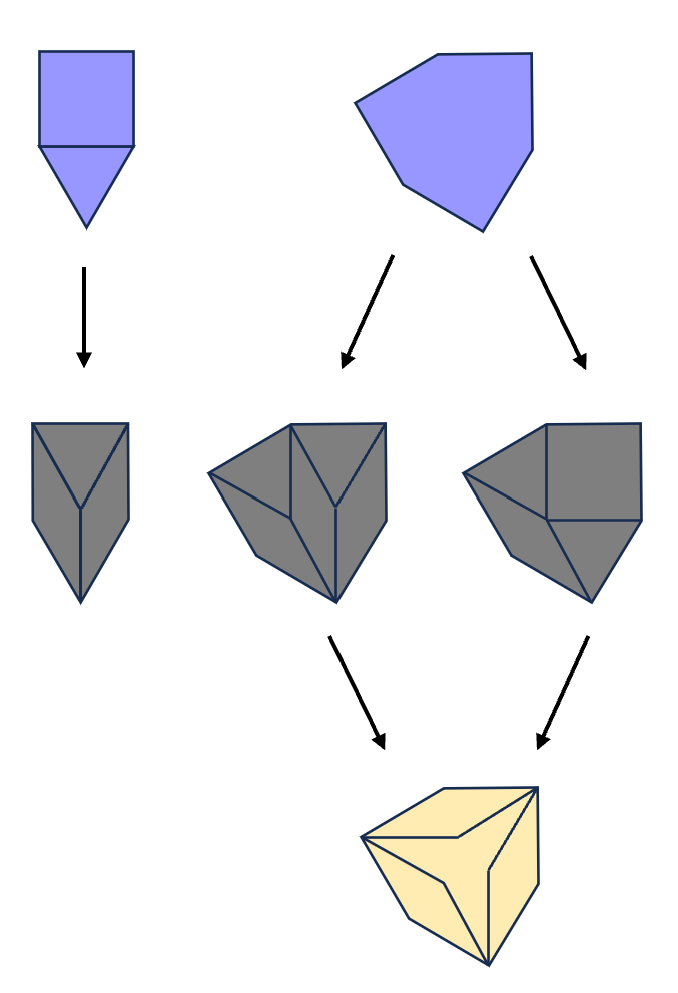}}
\caption{Local changes in tile configurations, associated with the addition of tiling vertices: 
(a) arising from the sequence of polygonal coincidence windows of increasing area in Fig.~\ref{fig:dod polygon windows}; 
(b) arising from replacing a polygonal coincidence window with its inscribing circle; 
(c) arising from the sequence of circular coincidence windows of increasing area in Fig.~\ref{fig:circ windows}.}
\label{fig:local changes}
\end{figure}

The resulting tilings are displayed in Figs.~\ref{fig:dod a}--\ref{fig:dod c}. As outlined in Fig.~\ref{fig:local changes a}, in going from the so-called \emph{shield tiling} of Fig.~\ref{fig:dod a} to the slightly denser tiling of Fig.~\ref{fig:dod b}, every shield is divided either into a square, a $30^\circ$-rhomb, and two triangles (by adding one vertex), or into two triangles and three rhombs (by adding two vertices). Continuing on from this tiling---identified here for convenience as the \emph{Niizeki--Gähler tiling}---to the densest of the three, shown in Fig.~\ref{fig:dod c}, some of the square--triangle configurations are redivided into a pair of rhombs and a single triangle by adding one vertex. We identify this third tiling as the \emph{Stampfli tiling}, as it was first described by Stampfli~\cite{Stampfli86}.

\begin{figure*}[ptbh]
\centering
\subfloat[]{\label{fig:circ windows} 
    \raisebox{0.045\textwidth}{\includegraphics[width=0.22\textwidth]{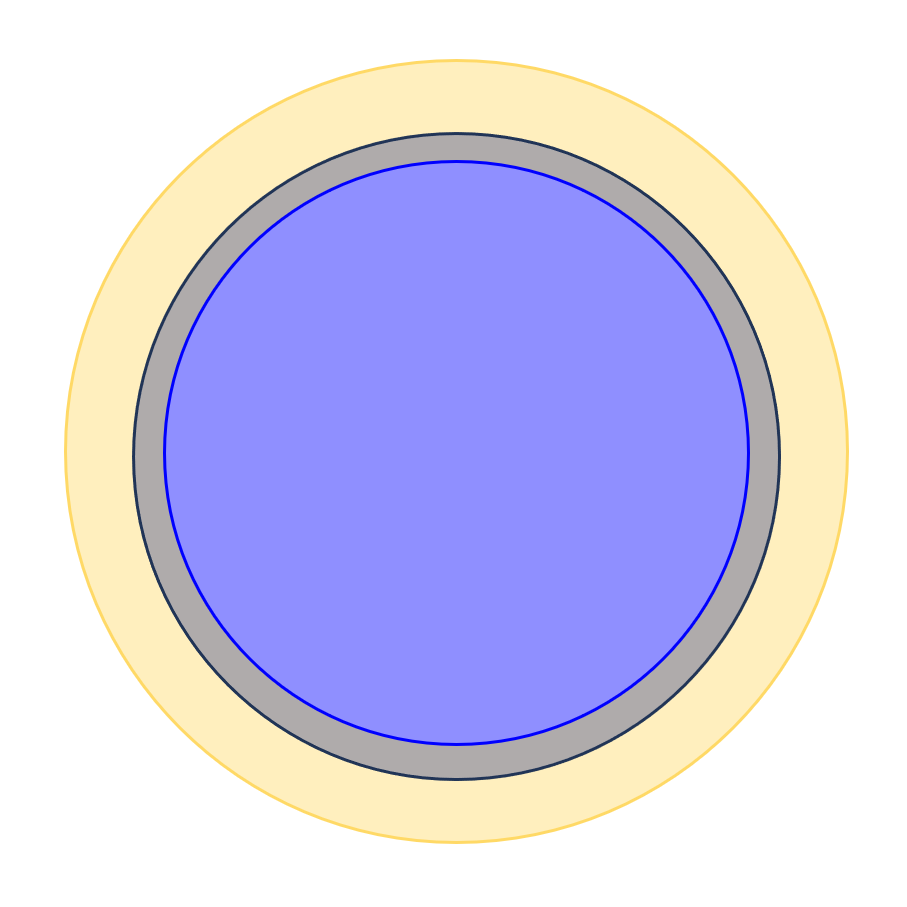}}}
\hfill
\subfloat[]{\label{fig:dod circ a} 
    \includegraphics[width=0.24\textwidth]{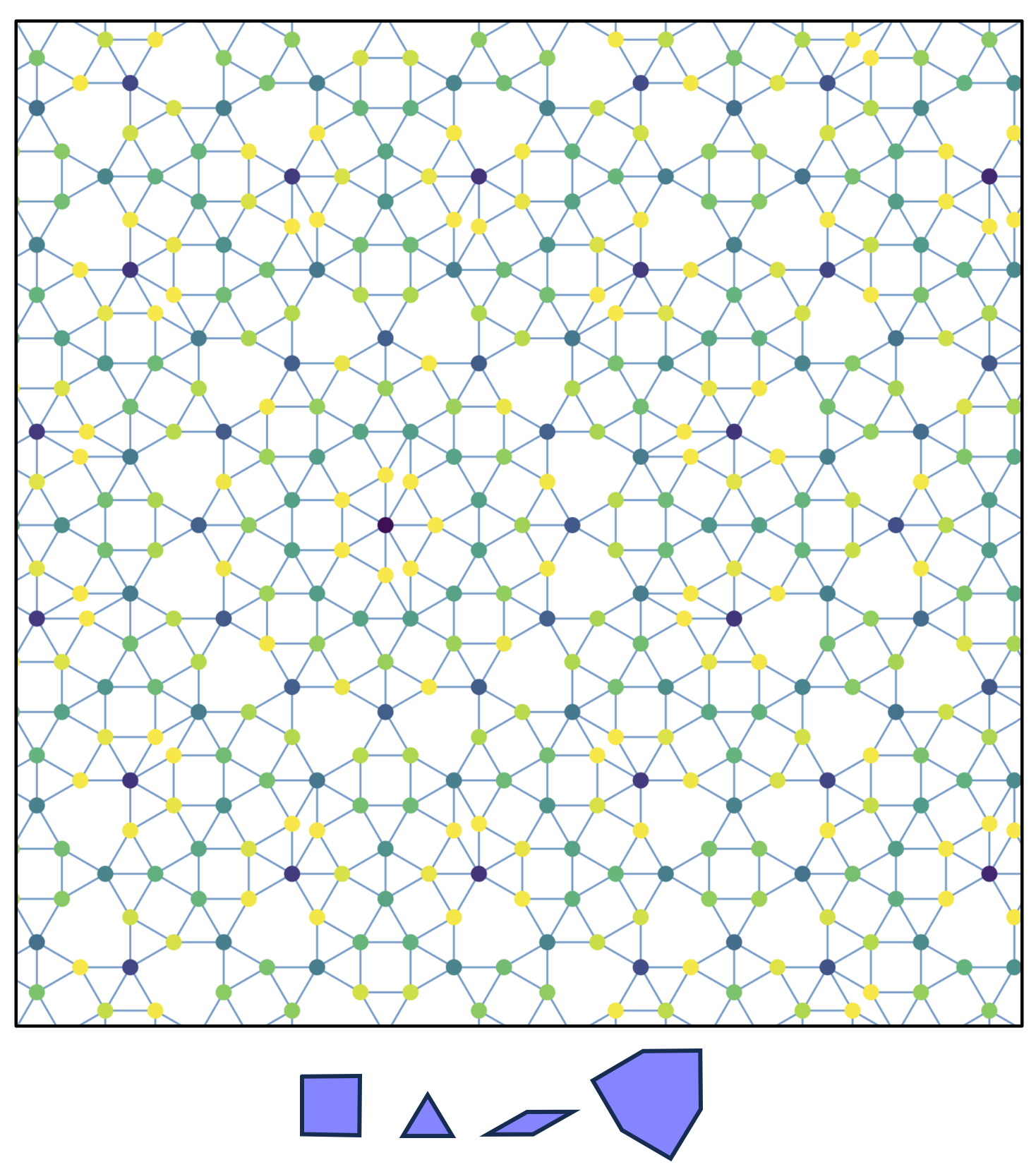}}
\hfill
\subfloat[]{\label{fig:dod circ b} 
    \includegraphics[width=0.24\textwidth]{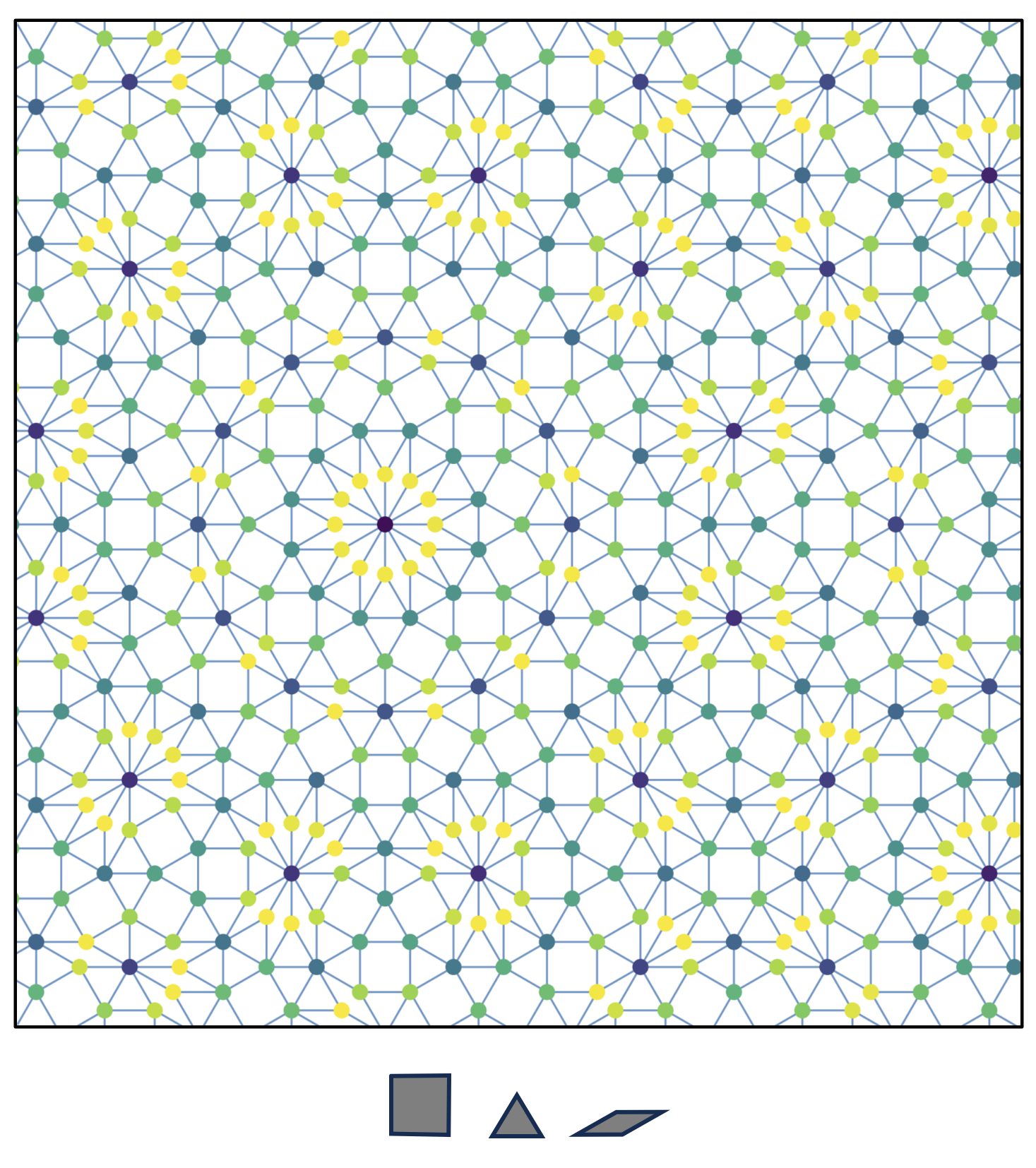}}
\hfill
\subfloat[]{\label{fig:dod circ c} 
    \includegraphics[width=0.24\textwidth]{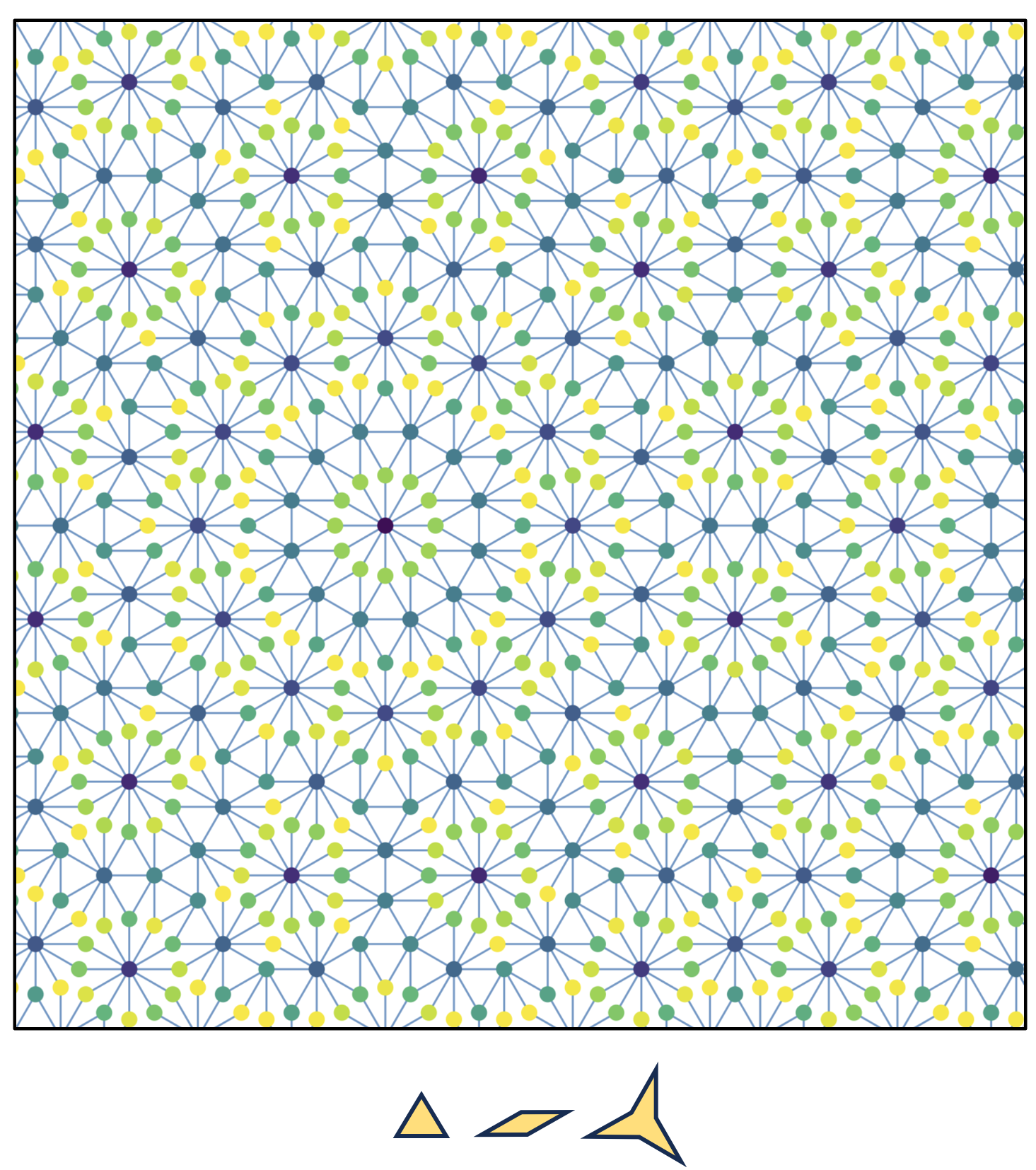}}
\caption{Three new dodecagonal tilings obtained by using circular coincidence windows.
(a) The three circular coincidence windows used for constructing the displayed tilings.
(b) A variant of the shield tiling, produced with the inscribing circle of the blue dodecagon of Fig.~\ref{fig:dod polygon windows}. 
(c) A variant of the Stampfli tiling, produced with the inscribing circle of the dark-gray dodecagon of Fig.~\ref{fig:dod polygon windows}. 
(d) The tiling generated with the large yellow circular coincidence window, containing three prototiles---a triangle, a rhomb, and a three-arm star, or \emph{tristar}---with no squares or shields.
}
\label{fig:dodecagonal with circle}
\end{figure*}

In the octagonal case we began with a circular coincidence window and refined it into an octagon in order to eliminate excess points. Here we begin with the known dodecagonal acceptance domains of the cut-and-project approach, and employ them directly as coincidence windows to reproduce well-known dodecagonal tilings. It is natural to consider replacing the polygonal windows by their inscribing circles and examining the effect of the new vertices that are added. These windows are shown as blue and dark-gray circles in Fig.~\ref{fig:circ windows} to emphasize their relative sizes. Because the added vertices are introduced isotropically along the boundary, the symmetry of the tiling is preserved. Moreover, because the difference in areas between a dodecagon and its inscribing circle is small, the density of vertices is only slightly increased. The added vertices do not require additional edge lengths, as they are connected to preexisting vertices at the correct distance, without introducing crossing edges. Thus, no clean-up is required, and these vertices may be included without compromising consistency or simplicity, in contrast to the octagonal case.

When replacing the small blue dodecagon with its inscribing circle, a fraction of the shields in the shield tiling (Fig.~\ref{fig:dod a}) acquire a single additional vertex, dividing them into a square, a rhomb, and a pair of triangles (Fig.~\ref{fig:local changes b}), while the rest remain intact. The new tiling, shown in Fig.~\ref{fig:dod circ a}, thus contains four prototiles, as the rhomb is added while the shield remains. The effect of replacing the large dark-gray dodecagon with its inscribing circle is to replace some of the square--triangle pairs in the Stampfli tiling (Fig.~\ref{fig:dod c}) with a pair of rhombs and a triangle. This increases the fraction of rhombic tiles.  

As a final observation, note that the tiling of Fig.~\ref{fig:dod circ b}, produced using the large dark-gray isotropic coincidence window, still contains many square--triangle pairs that can accommodate an additional vertex. Similarly, all the original shields contain only one or two additional vertices, while there is clearly a third position in each that can accommodate another without violating the single-edge-length preference. Indeed, there exists a range of coincidence thresholds $r$ that increases the density just enough to add all these vertices---but not more---thus producing a tiling without squares, in which all shields are replaced by three rhombs surrounding a 3-fold nonconvex star, which we refer to as the three-arm star, or \emph{tristar} tile (see Fig.~\ref{fig:local changes c}). The resulting tiling, generated with the large yellow coincidence window of Fig.~\ref{fig:circ windows}, is shown in Fig.~\ref{fig:dod circ c}. We note that the same three-arm star tile was observed independently by Sadoc and Imp\'{e}ror-Clerc~\cite{Sadoc23} in a different dodecagonal tiling that still contains square tiles.

\section{Fibonacci tilings by near coincidence} 
\label{Sec:Fibonacci}

In the previous sections we focused on twisted bilayers with identical length scales, generating both well-known tilings and new variants by means of the near-coincidence method. We now turn to the role of the scaling factor, without introducing any relative rotation, and use the Fibonacci tiling family as our case study. We begin with the one-dimensional Fibonacci tiling~\cite{Baake24,Jagannathan21}, which captures the essential features of the construction, and then show how it extends naturally to two-dimensional square~\cite{Lifshitz02} and hexagonal~\cite{Coates24} Fibonacci tilings. Both examples arise from a scaled bilayer construction, in which one layer is rescaled relative to the other by the golden mean $\tau=(1+\sqrt{5})/2$.

\subsection{One-dimensional Fibonacci tilings from scaled bilayers}

To expose the essential features of the scaled bilayer construction, we first consider the one-dimensional case. The cut-and-project construction of the one-dimensional Fibonacci tiling may be formulated as a projection of the square lattice $\mathbb{Z}^2$, with physical-space coordinates $(n_0 + n_1\tau)$ and internal-space coordinates $(n_0\tau - n_1)$. The acceptance condition selects those integer pairs $(n_0,n_1)$ for which
\begin{equation}
    n_0\tau - n_1 \in \text{AD},
\end{equation}
with a canonical acceptance domain whose length $|\text{AD}|=1+\tau=\tau^2$, yielding the standard Fibonacci tiling with edge lengths in the ratio $\tau\!:\!1$.

Within the near-coincidence framework, the same integer pairs are selected by viewing the acceptance domain AD as a coincidence window CW of the same size. However, the tiling vertices are placed at the midpoint positions $(n_0\tau + n_1)/2$, rather than at $(n_0 + n_1\tau)$. As a result, one obtains the correct Fibonacci ordering of long and short edges, but not their canonical length ratio: the “short” and “long” edges initially appear with lengths $\tau/2$ and $1/2$, respectively.

\begin{figure}[tb]
\centering
\begin{tikzpicture}
    \node[anchor=south west, inner sep=0] (base) at (0,0)
        {\includegraphics[width=0.95\columnwidth]{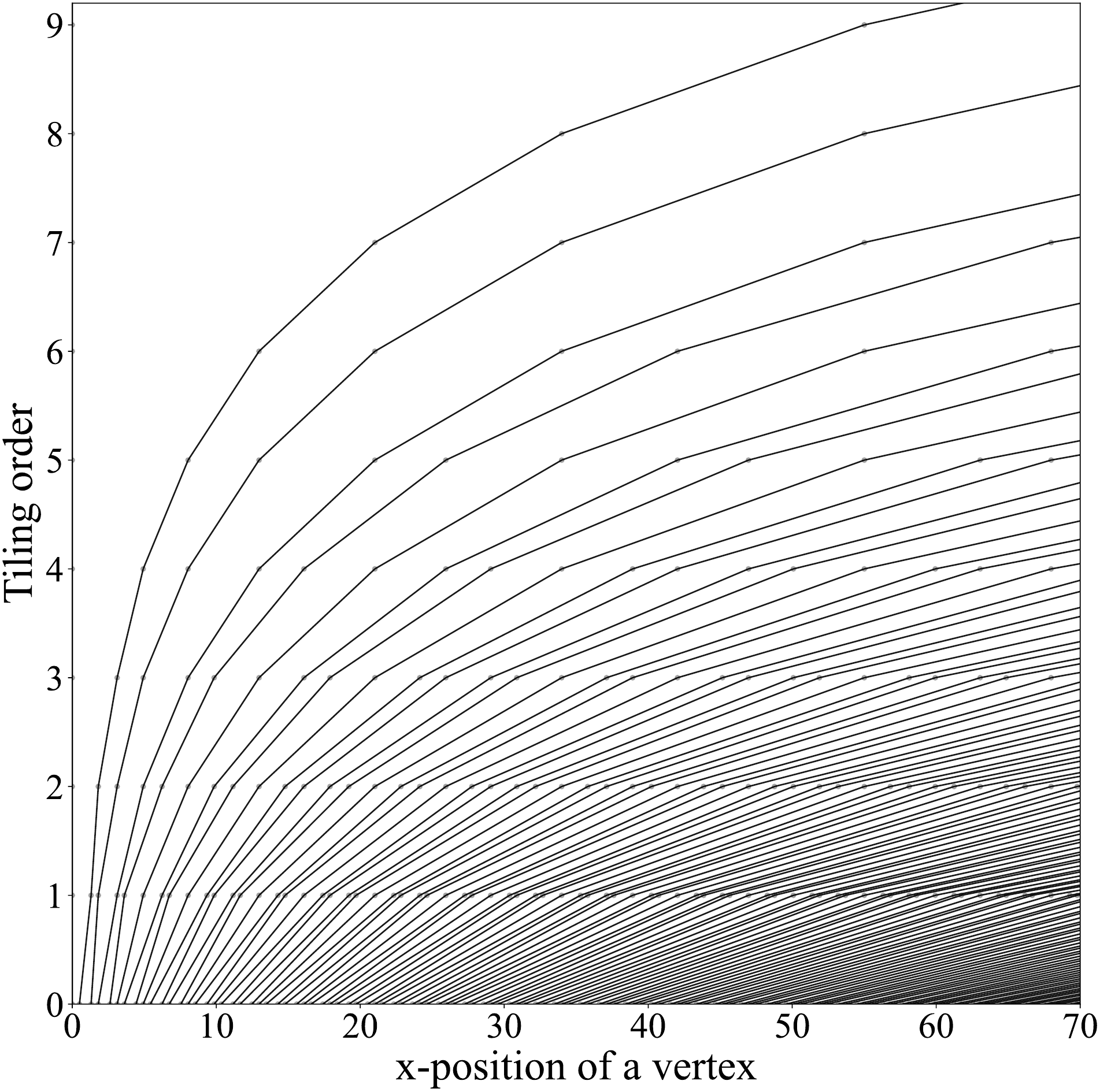}};

    \node[anchor=south west, inner sep=0, opacity=0.9] at (0.08\columnwidth,0.62\columnwidth)
        {\includegraphics[width=0.23\columnwidth]{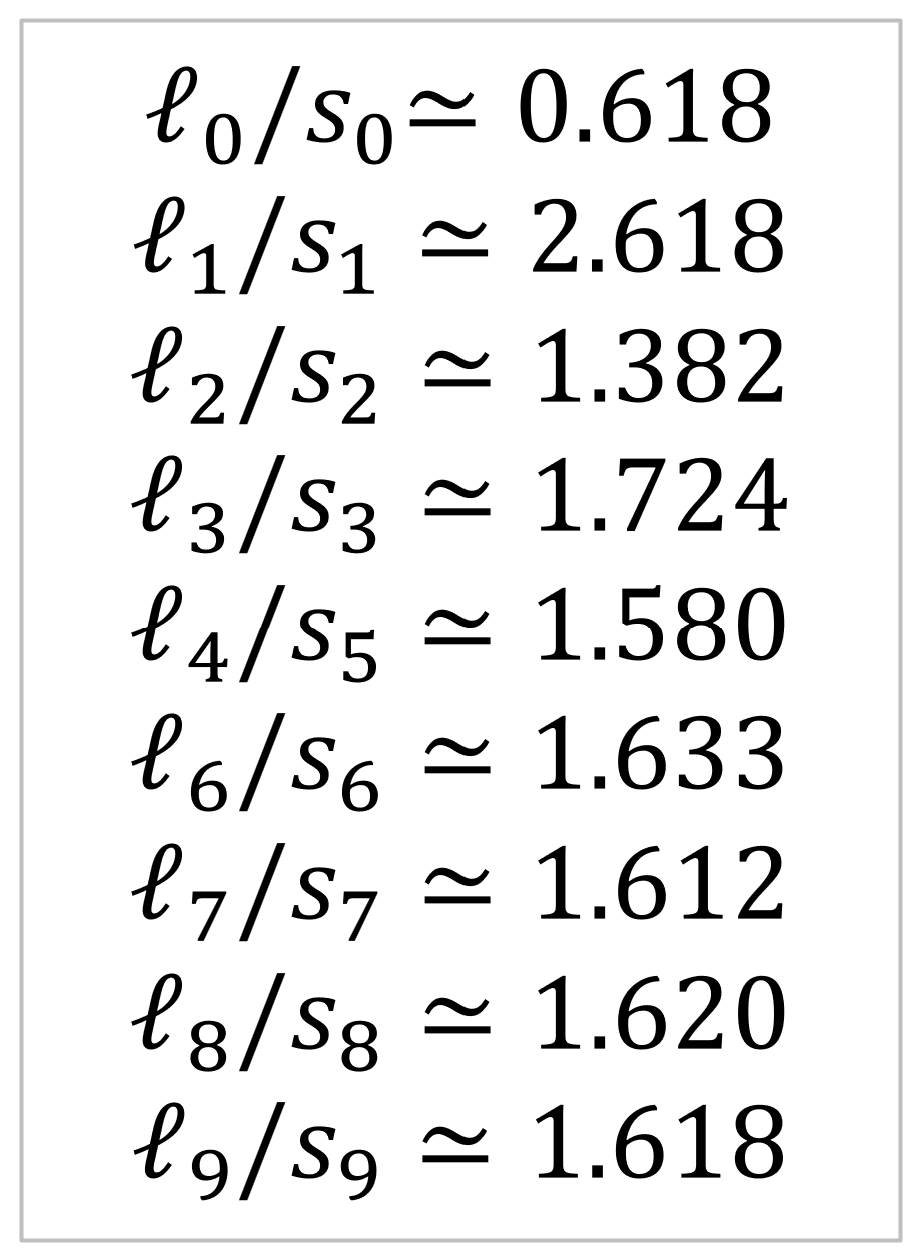}};

    \node[anchor=south west, inner sep=0, opacity=0.9] at (0.48\columnwidth, 0.58\columnwidth)
        {\includegraphics[width=0.45\columnwidth]{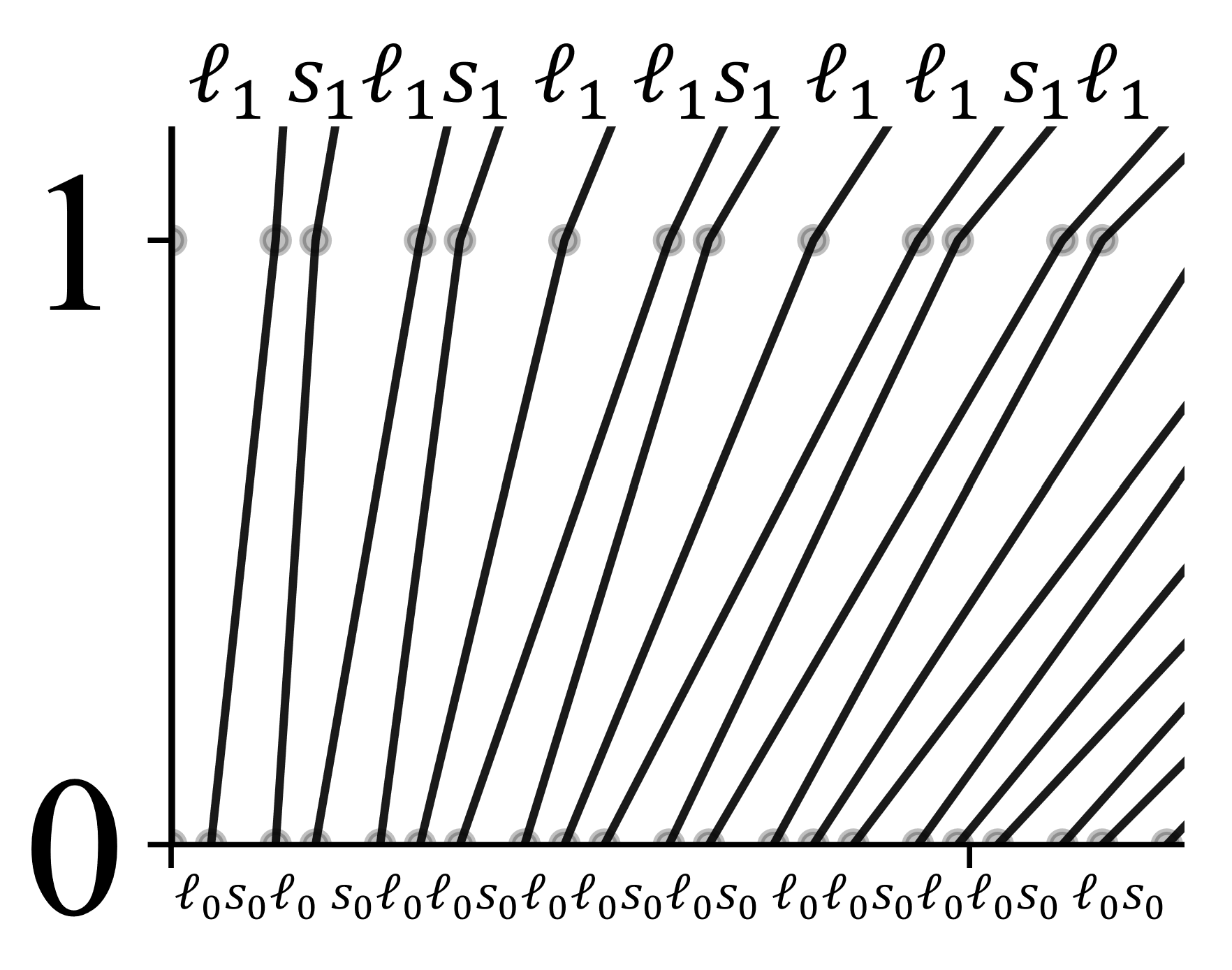}};
\end{tikzpicture}
\caption{Successive one-dimensional Fibonacci tilings for centered coincidence windows of diminishing size $\tau^{2-n}$, shown row by row. The correct Fibonacci order is obtained for all $n$, while the edge-length ratio $l_n/s_n$ approaches $\tau$ only with increasing order, as detailed in the upper-left inset. The upper-right inset demonstrates that at order $n=0$, with the largest coincidence window, one begins with the inverted ratio relative to the canonical one, where $l_0$ labels short edges and $s_0$ labels long ones.}
\label{fig:1d fib}
\end{figure}

This discrepancy can be resolved in two ways. One may simply retain the correct sequence of long and short edges and rescale their lengths by hand to enforce the canonical ratio. Alternatively, one may achieve any required accuracy by shrinking the coincidence window by successive factors of $\tau$, thereby generating sparser tilings whose edge-length ratio approaches $\tau$. This occurs because $\tau$ is a Pisot number---an eigenvalue greater than unity of a substitution matrix whose remaining eigenvalue $-1/\tau$ is less than unity in magnitude---ensuring that successive substitutions drive the edge-length ratio toward its canonical value, even when starting from the inverted ratio. This behavior is illustrated in Fig.~\ref{fig:1d fib}, where successive coincidence windows of diminishing size are used.

A direct attempt to recover the canonical Fibonacci substitution rules: $S \rightarrow L$ and $L \rightarrow LS$ by overlaying a pair of $\tau$-scaled tilings, fails as well. Instead, one obtains alternative \emph{symmetric} substitution rules~\footnote{Alternative, or conjugate, substitution rules that define the same tiling are discussed elsewhere. See, for example, the discussion of metallic-mean sequences in Baake and Grimm~\cite[Sec.~4.4]{Baake13}.} in which every $LL$ pair is replaced by $SLLS$, and isolated $L$ tiles are substituted by either $LS$ or $SL$, with equal probability, depending on their surroundings. This can be seen by comparing any pair of successive one-dimensional tilings in Fig.~\ref{fig:1d fib}. These rules produce the same one-dimensional Fibonacci tiling as the canonical ones, but require keeping track of tile orientations. This can be achieved by introducing oriented tiles using overhead arrows, and removing the arrows at the end, with substitution rules given by $\LR \to \LR\SR$, $\LL \to \SL\LL$, $\SR \to \LL$, and $\SL \to \LR$. Compare with the use of red arrows in Fig.~\ref{fig:Substitution rules} for substitution rules of the Ammann-Beenker tiling.

This behavior arises from the use of a \emph{symmetric} coincidence window, in which the difference vector is measured with respect to the center of the window. If instead one measures the difference vector with respect to the left edge of the window---thereby breaking the symmetry between $\red{\mathbf{p}_1}$ and $\blue{\mathbf{p}_2}$ by admitting nearly coincident blue-red pairs while rejecting nearly coincident red-blue pairs---one recovers the canonical substitution rules, which globally replace $L$ by either $LS$ or $SL$. However, this global choice alternates between successive steps, effectively flipping the orientation of tiles at each iteration, which can be captured by the following oriented-tile substitution rules: 
$\LR \to \LL\SL$, $\LL \to \SR\LR$, $\SR \to \LL$, and $\SL \to \LR$.

\subsection{Square Fibonacci tiling}

\begin{figure}[b]
\centering
\subfloat[]{\label{fig:square a} \fbox{\includegraphics[width=0.3\linewidth]{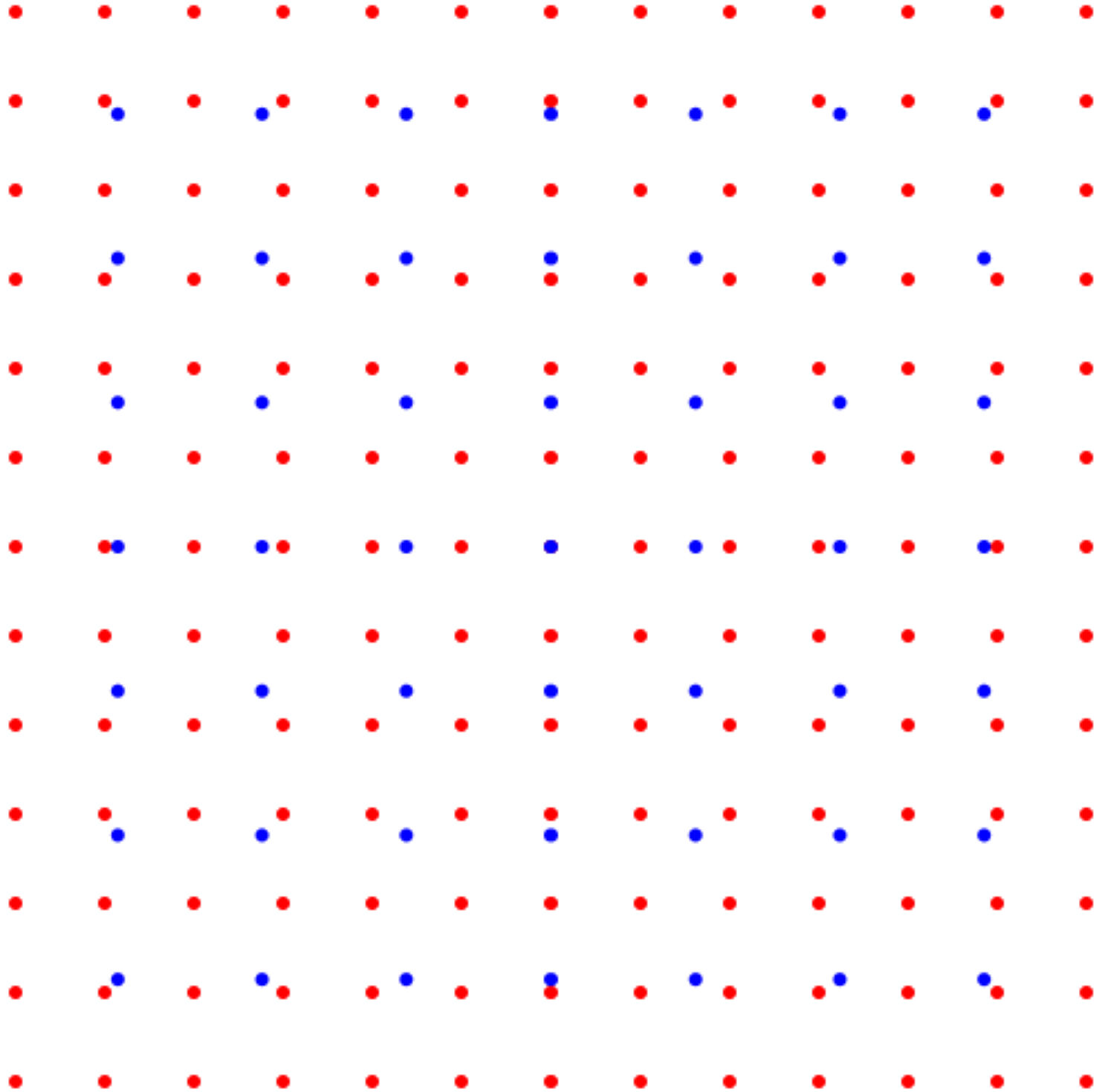}}}
\hfill
\subfloat[]{\label{fig:square b} \fbox{\includegraphics[width=0.3\linewidth]{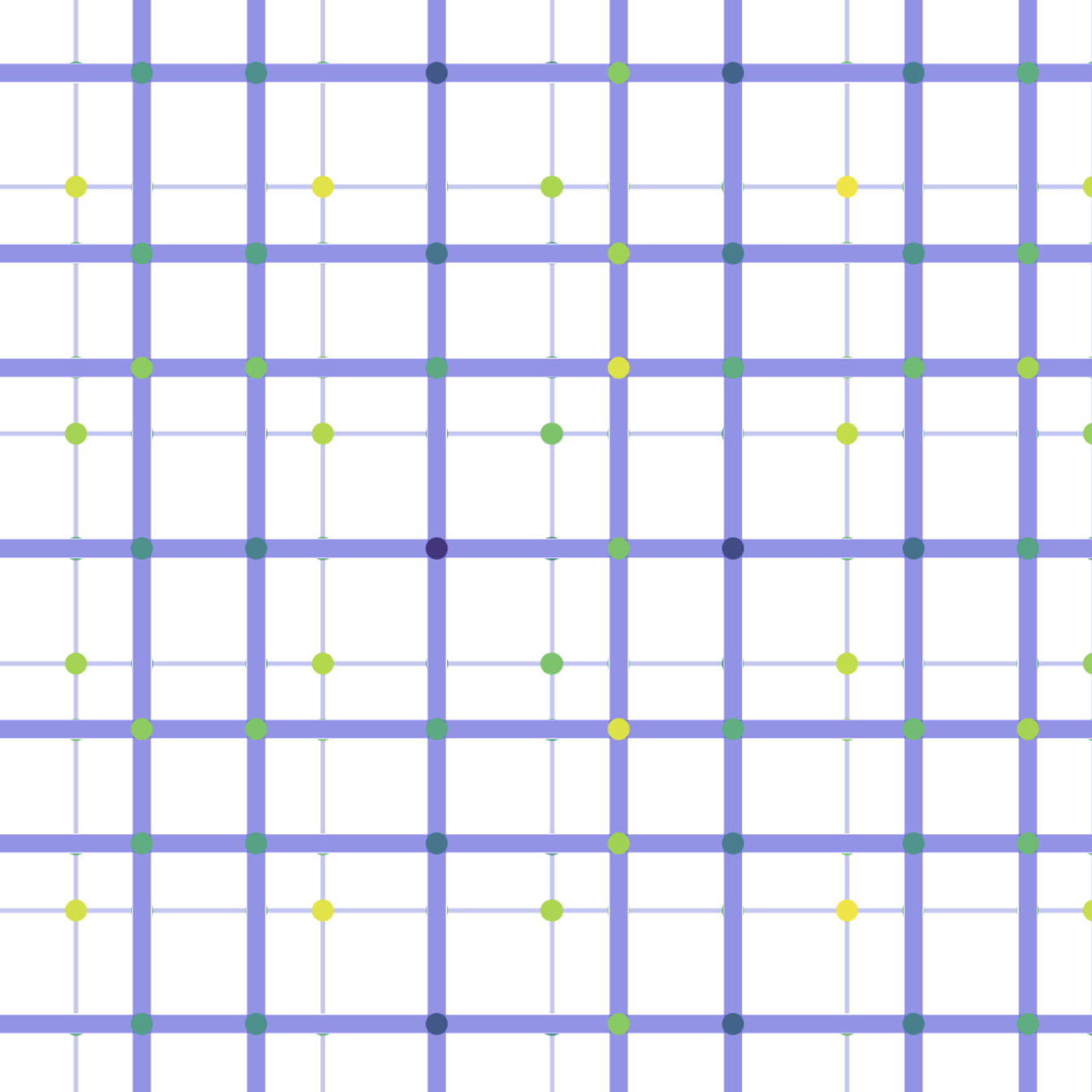}}}
\hfill
\subfloat[]{\label{fig:square c} \fbox{\includegraphics[width=0.3\linewidth]{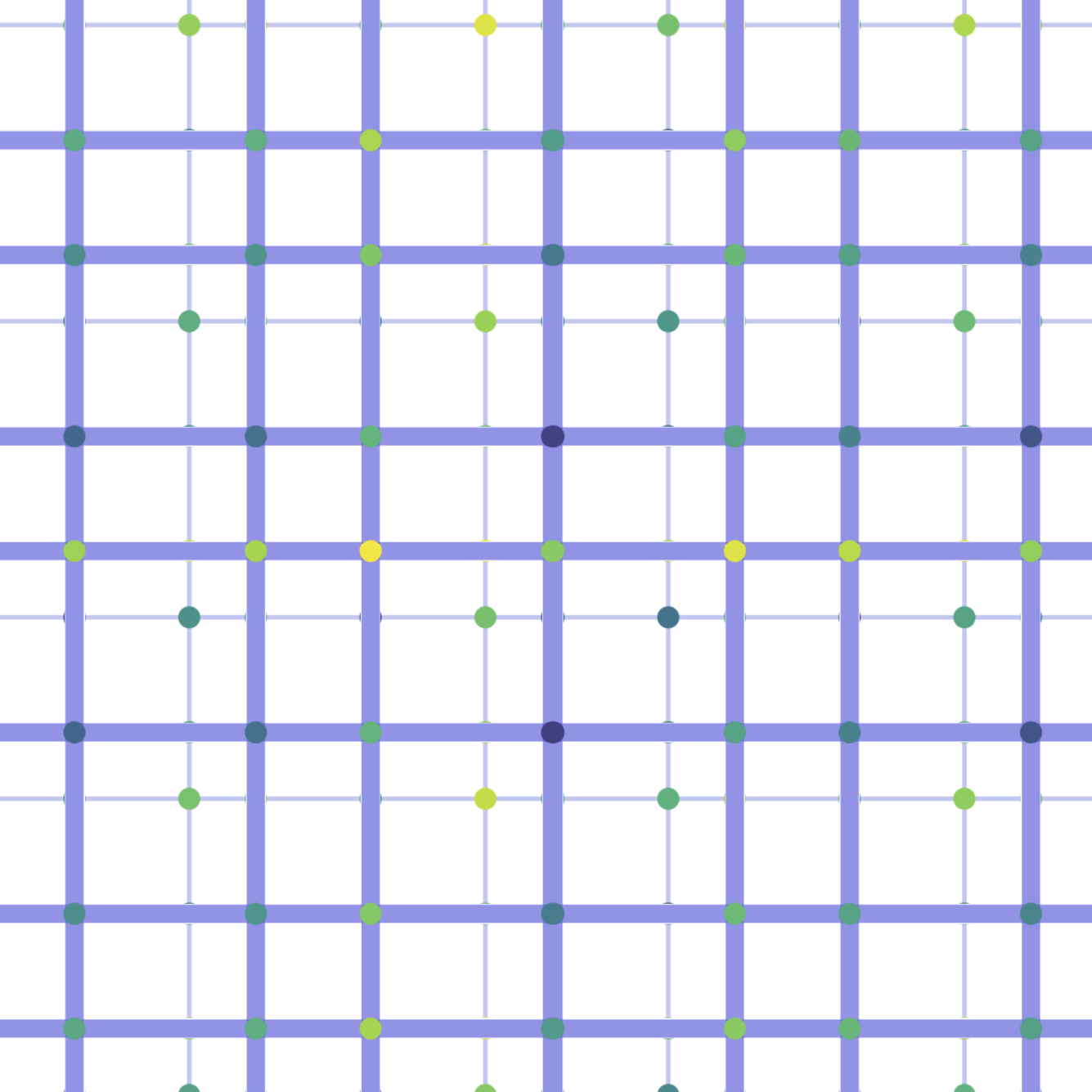}}}
\caption{Square Fibonacci tilings obtained from a $\tau$-scaled bilayer: (a) the scaled layers; (b) a pair of tilings, obtained using centered $\tau$-scaled coincidence windows, which lead to symmetric substitution rules, revealing all four orientations of the canonical substitution rules~\cite[Fig.~3]{Lifshitz02}; (c) using shifted coincidence windows yields the canonical Fibonacci substitution with alternating orientation.}
\label{fig:Square fibonacci}
\end{figure}

The two-dimensional square Fibonacci tiling~\cite{Lifshitz02} follows immediately as a direct product extension of the one-dimensional construction. Using a $\tau$-scaled square bilayer (Fig.~\ref{fig:square a}) and a square coincidence window, one obtains the tiling shown in Fig.~\ref{fig:square b}. The tiling consists of three prototiles: a small $S\times S$ square, a large $L\times L$ square, and an $S\times L$ rectangle, with the Fibonacci sequence appearing independently along both the horizontal and the vertical directions.

\begin{figure}[tb]
\centering
\subfloat[]{\label{fig:star sq} \includegraphics[width=0.48\columnwidth]{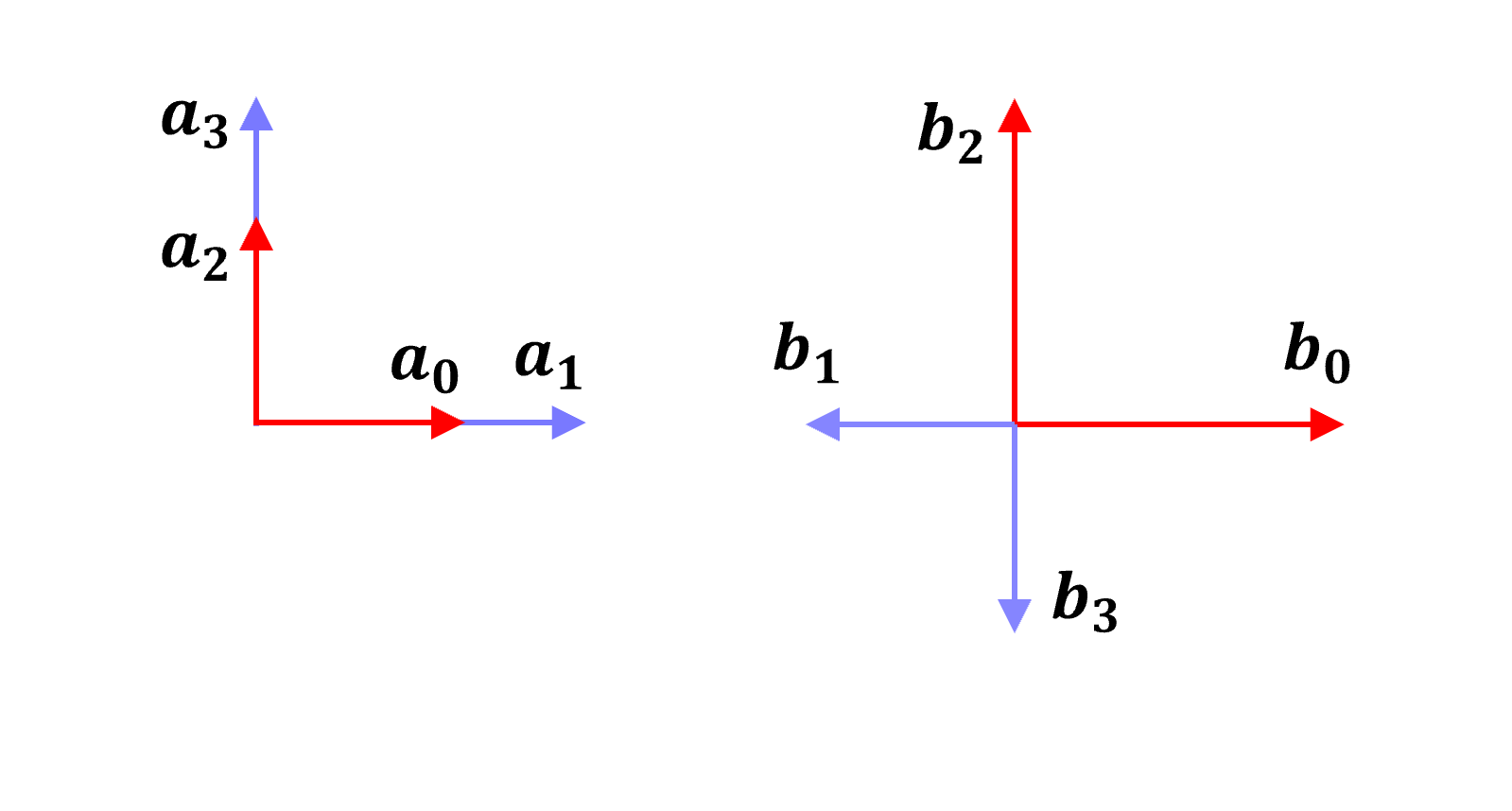}}
\hfill
\subfloat[]{\label{fig:star hex} \includegraphics[width=0.48\columnwidth]{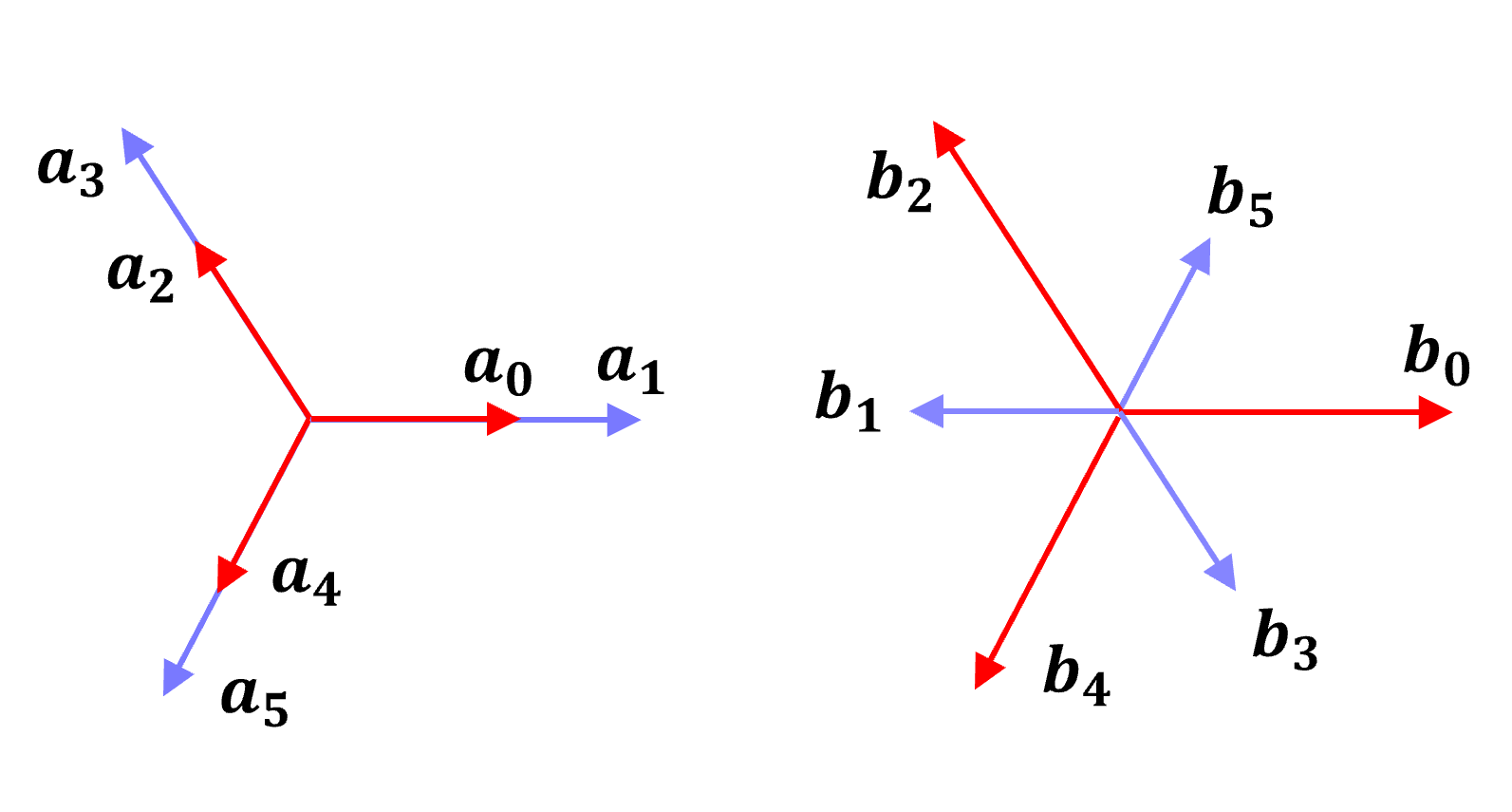}}
\caption{The standard star maps used in the cut-and-project method for the (a) square; and (b) hexagonal Fibonacci tilings.}
\label{fig:Star-map fib}
\end{figure}

The corresponding cut-and-project star-map, shown in Fig.~\ref{fig:star sq}, as well as the distinction between centered and shifted coincidence windows, carry over directly from the one-dimensional case. Here as well, one observes the same discrepancy where the edge-length ratio approaches its canonical value only as the coincidence window shrinks; and centered coincidence windows lead to symmetric substitution rules as shown in Fig.~\ref{fig:square b}, while shifted windows recover the canonical Fibonacci substitution, as illustrated in Fig.~\ref{fig:square c}.

\subsection{Hexagonal Fibonacci tiling}

\begin{figure}[tb]
\centering
\subfloat[]{\label{fig:hexagonal a} \includegraphics[width=0.45\columnwidth]{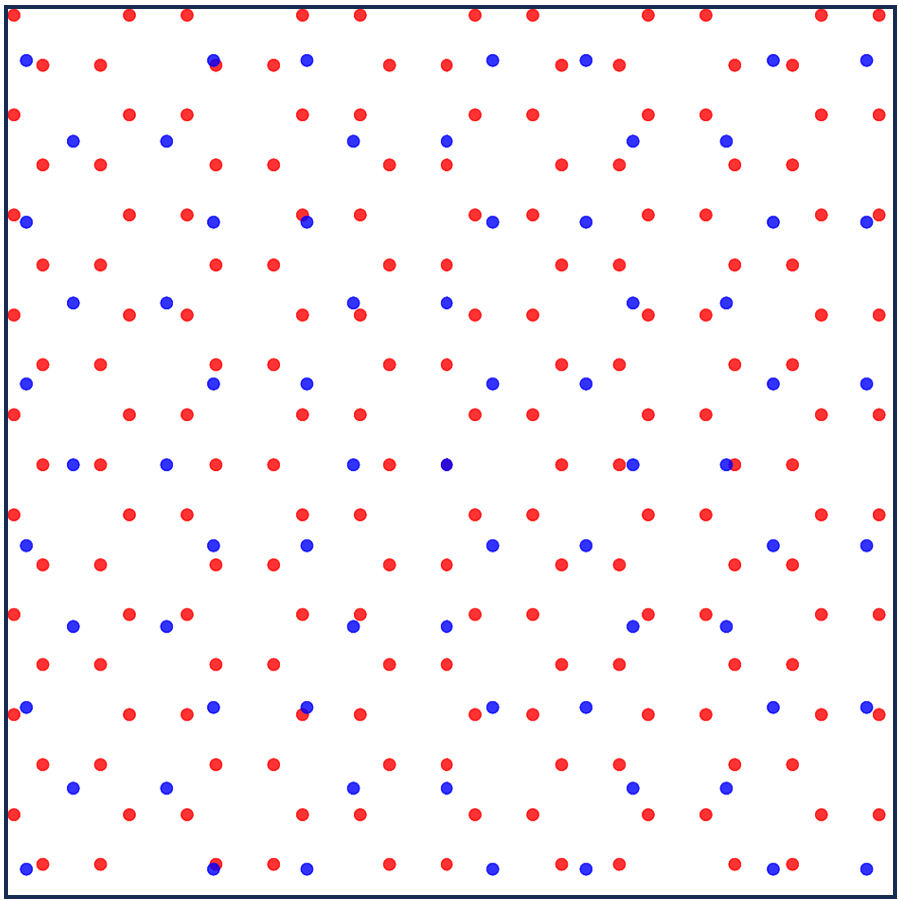}}
\hfill
\subfloat[]{\label{fig:hexagonal b} \includegraphics[width=0.45\columnwidth]{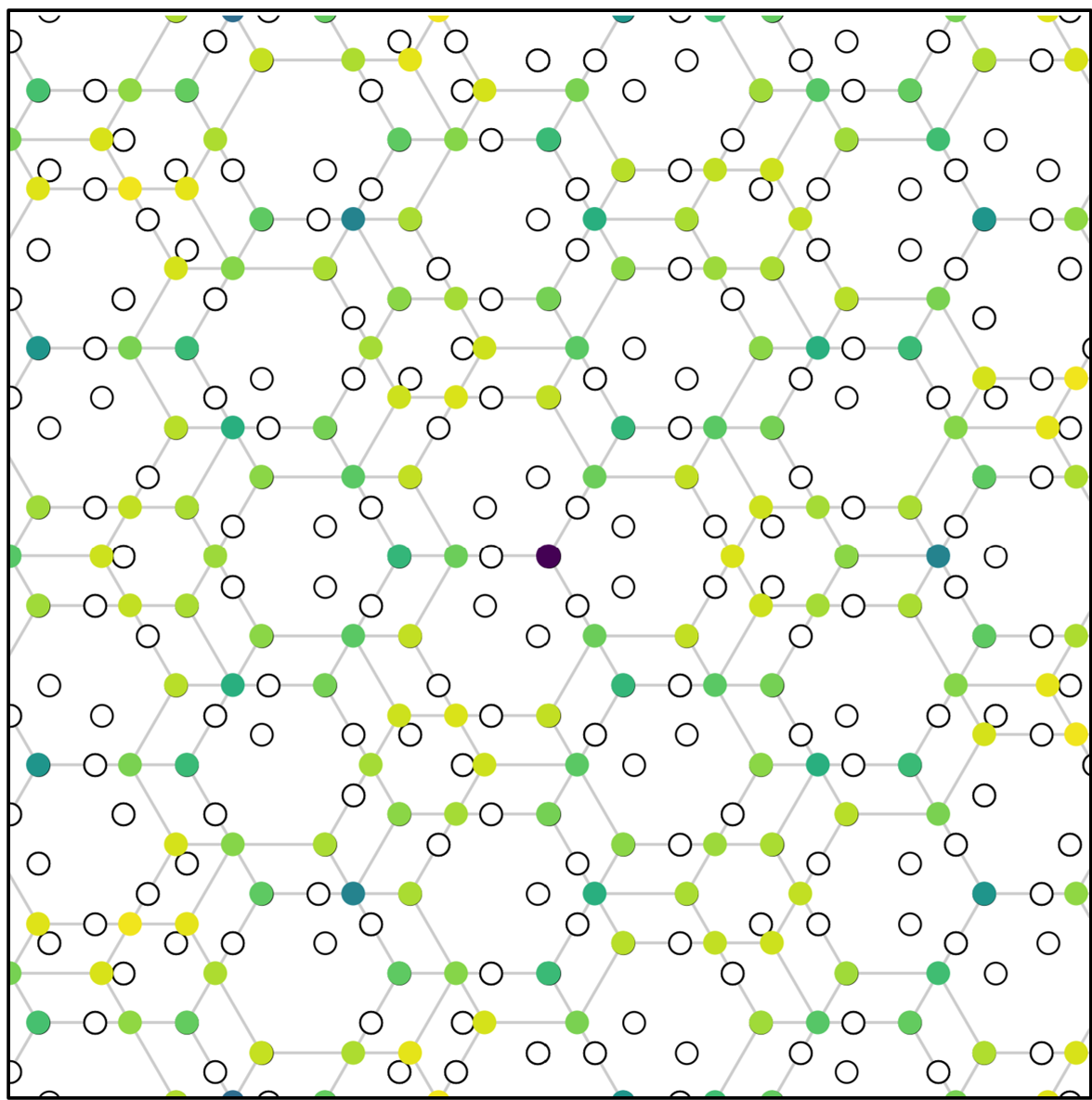}}
\hfill
\subfloat[]{\label{fig:hexagonal c} \includegraphics[width=0.45\columnwidth]{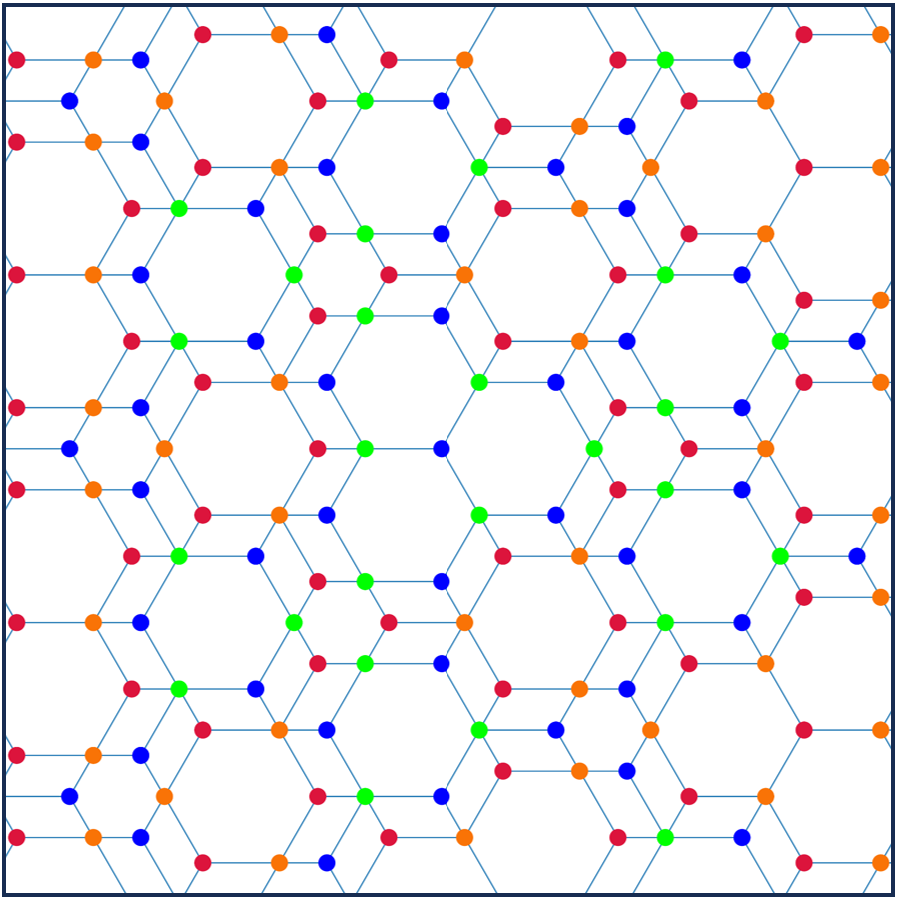}}
\hfill
\subfloat[]{\label{fig:hexagonal d} \includegraphics[width=0.45\columnwidth]{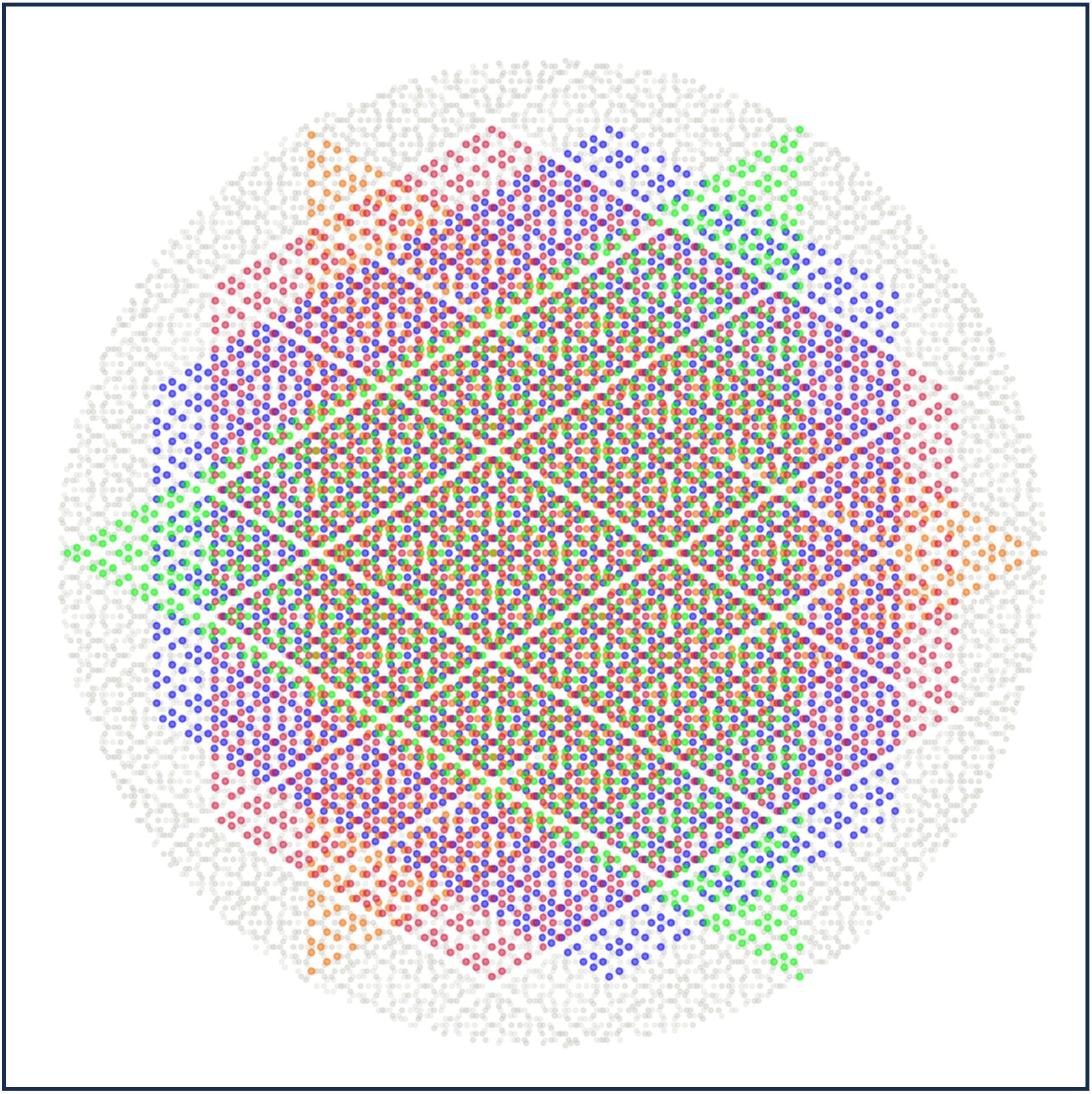}}
\caption{The $H_{00}$ Hexagonal Fibonacci tiling~\cite{Coates24}.
(a) A $\tau$-scaled honeycomb bilayer. 
(b) The hexagonal Fibonacci tiling, with excess points marked by empty circles. 
(c) The tiling after removing the excess points, with vertices color-coded according to their parity classes, as described in the text. 
(d) The four coincidence windows corresponding to the four vertex types, with rejected points shown in gray.}
\label{fig:Hexagonal fibonacci}
\end{figure}

We next consider the near coincidence of a $\tau$-scaled honeycomb bilayer, shown in Fig.~\ref{fig:hexagonal a}. In this case the optimal shape of the coincidence window is not obvious. Nevertheless, a natural default choice is to use a sufficiently small circular coincidence window, ensuring (to any desired accuracy) the canonical edge-length ratio, as discussed in the previous section, without affecting the hexagonal symmetry. This produces the tiling in Fig.~\ref{fig:hexagonal b}, containing excess points, marked by empty circles, that are easily removed by simple local rules: eliminating \emph{dangling vertices} that connect to only one other vertex, and \emph{redundant vertices} that lie along existing edges. The resulting clean tiling, shown in Fig.~\ref{fig:hexagonal c}, is precisely the $H_{00}$ hexagonal Fibonacci tiling of Coates \emph{et al.}~\cite{Coates24}.

Mapping the accepted and rejected vertices onto the circular coincidence window, shown in Fig.~\ref{fig:hexagonal d}, reveals a central hexagon containing only accepted vertices, surrounded by various mixed regions containing both accepted and rejected vertices, which in turn are surrounded by an outer perimeter containing only rejected vertices. Further inspection suggests that four coincidence windows are required, corresponding to four distinct subsets of vertices. These depend on the parities of the numbers of small red steps and large blue steps---indicated in Fig.~\ref{fig:star hex}---required to reach a vertex from the origin. This is in agreement with the cut-and-project construction of this tiling, which requires four different two-dimensional cuts through a six-dimensional embedding space (for details see Ref.~\cite{Coates24}).

The tiling in Fig.~\ref{fig:hexagonal c} is colored according to this scheme, and so are the corresponding points mapped onto the coincidence window in Fig.~\ref{fig:hexagonal d}. This partitioning of the vertices yields a pair of oppositely oriented orange and green triangular coincidence windows, and a pair of oppositely oriented red and blue 3-fold symmetric hexagonal coincidence windows, thus maintaining the overall hexagonal symmetry.

Thus, in contrast to the square case, the hexagonal Fibonacci tiling involves a richer structure, with multiple vertex classes and multiple coincidence windows, reflecting the higher-dimensional nature of the corresponding cut-and-project construction.

\section{Summary and future directions}  
\label{Sec:Future}

Inspired by physical bilayer systems and their moiré patterns, we have introduced a new method for generating quasiperiodic tilings of the plane. The method is both intuitive and rigorous: intuitive because it builds directly on the near coincidence of points from superimposed periodic layers, and rigorous because it can be mapped, in a variety of ways, to the established cut-and-project formalism. Using this framework we have reproduced several well-known quasiperiodic tilings, including the octagonal Ammann--Beenker, the dodecagonal Niizeki--Gähler, and the square and hexagonal Fibonacci tilings, the latter obtained by first analyzing the underlying one-dimensional construction. Along the way, we have identified different ways in which the near-coincidence method corresponds to the cut-and-project approach: in some cases directly, in others after rescaling the coincidence window, and in the hexagonal Fibonacci case after recognizing the existence of several coincidence windows associated with distinct vertex classes. We note that other bilayer construction schemes were proposed over the years for generating quasiperiodic tilings, which bear some similarities with our near-coincidence method~\cite{Sadoc90,Stampfli12,Sadoc23}.

While the cut-and-project method remains far more powerful and general, the near-coincidence construction offers a perspective in which certain choices become natural that are rarely considered~\cite{Masakova03} in the cut-and-project framework. Chief among these is the use of a circular coincidence window---the simplest and most intuitive choice in our method, but one that is not normally employed in cut-and-project constructions. In this way we have discovered new quasiperiodic tilings that would not likely have been encountered otherwise, like the tiling of Fig.~\ref{fig:dod circ c}, containing the tristar tile~\footnote{A similar tristar tiling is obtained by using a larger dodecagonal coincidence window. See the last row in the triangular bilayer section of Tab.~\ref{tab:tiling parameters}.}.

An important observation concerns the diffraction spectrum, or equivalently the Fourier transform, of the near-coincidence tiling. In this framework, the bilayer diffracts as the superposition of the diffraction patterns of the rotated or rescaled individual layers, producing a mixed diffraction pattern that does not constitute a reciprocal lattice, or $\mathbb{Z}$-module, in the strict mathematical sense of closure under vector addition. However, the process of accepting only nearly coincident pairs of points and shifting them to their midpoints, which in physical systems may arise from interactions between the layers, generates additional harmonics at linear combinations of Bragg peaks from the separate layers. This gives rise to a well-behaved reciprocal lattice, exhibiting closure under vector addition along with the prescribed $N$-fold symmetry.

A practical advantage of the near-coincidence method lies in its algorithmic simplicity. Once the layers are defined, the construction requires only the specification of a coincidence threshold and the testing of near coincidences. This lends itself naturally to numerical implementation. Indeed, we have developed a web-based application, implemented by Ochana~\cite{Ochana25}, that accepts as input the choice of layers, their twist or scaling, and the desired coincidence window---circular or polygonal---and generates the corresponding quasiperiodic tiling. This tool has already proved invaluable for exploring known examples and for discovering new ones. All parameters for the tilings discussed here are provided in Appendix~\ref{sec:app-data}, and can be used directly within the application, presented in Appendix~\ref{sec:app-fib}.

\begin{figure}[tb]
\centering
\subfloat[]{\label{fig:dod 12-fold 3 layers} \includegraphics[width=0.48\columnwidth]{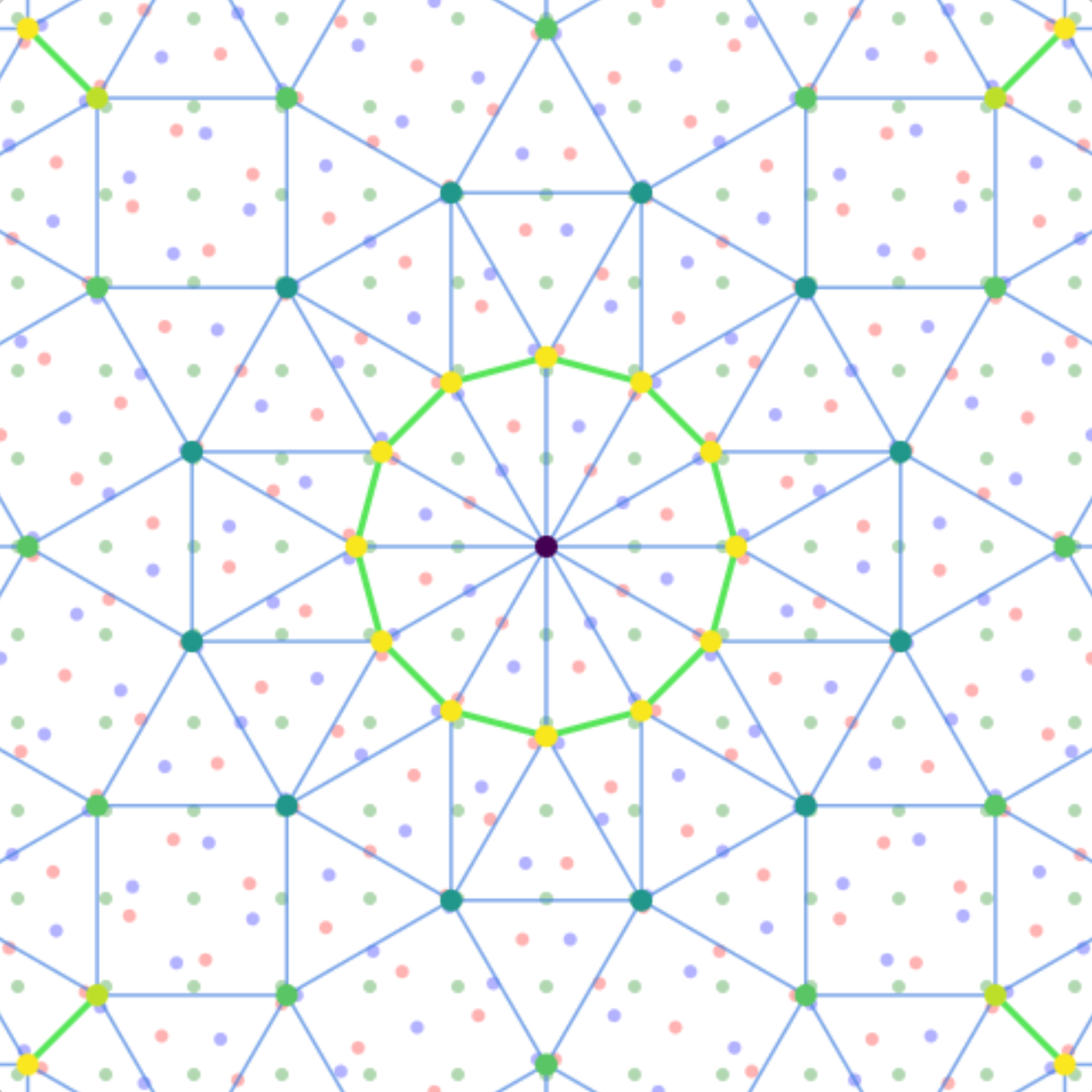}}
\hfill
\subfloat[]{\label{fig:dod 4-fold 2 layers} \includegraphics[width=0.48\columnwidth]{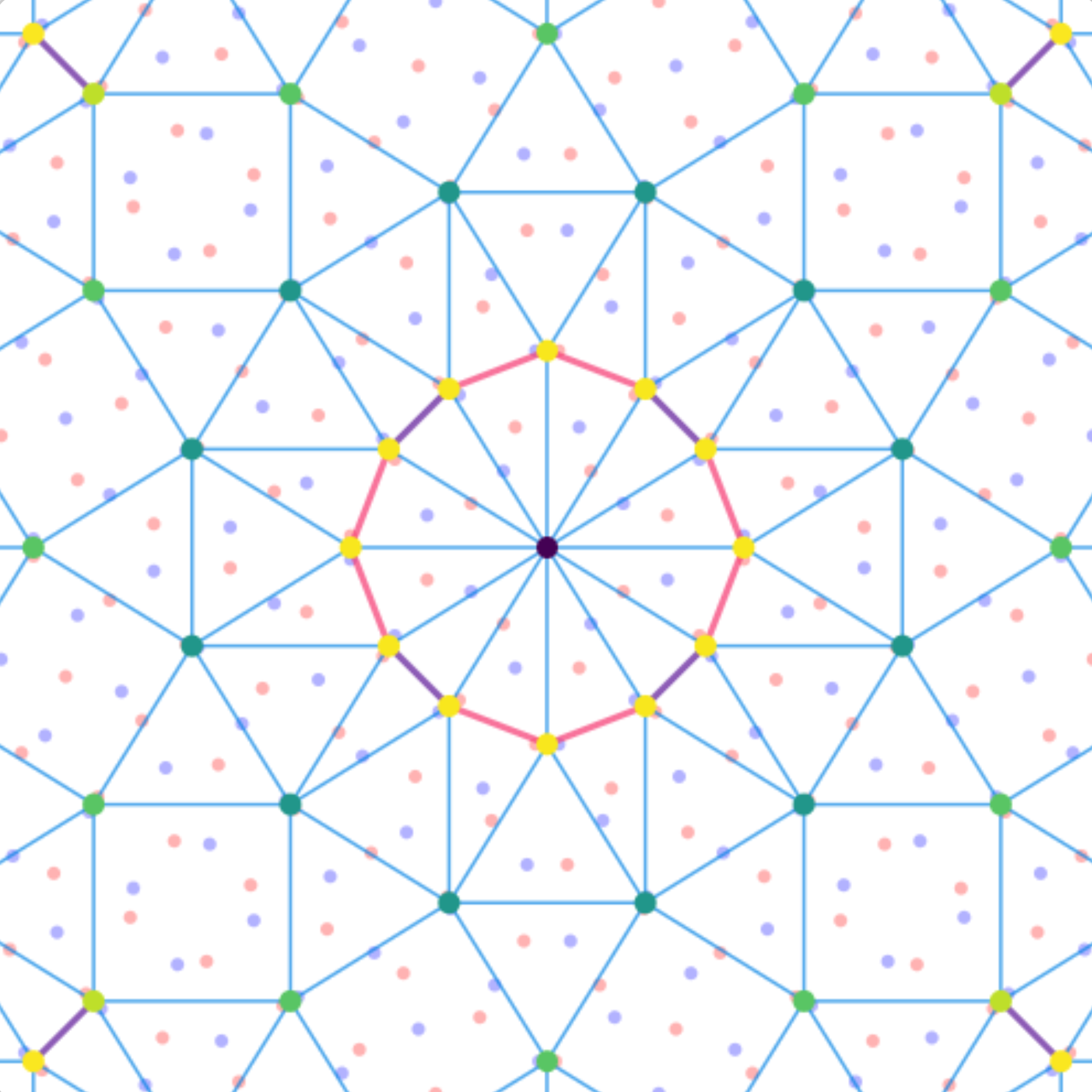}}
\caption{Two tilings produced via the near-coincidence method:
(a) a dodecagonal tiling generated from three square lattices---red, blue, and green---rotated by $30^\circ$ relative to one another; and
(b) a pseudo-dodecagonal tiling derived from only two square lattices---red and blue---rotated by $30^\circ$, possessing only tetragonal symmetry, as can be seen by the appearance of distorted rhombs, with two distinct short-diagonal lengths (highlighted in purple and red).  
}
\label{fig:dod tilings 2 layers}
\end{figure}

Looking ahead, several directions for future work suggest themselves. First, we intend to generalize the near-coincidence construction to multilayer systems. We have preliminary results showing that a dodecagonal tiling can be obtained from three square layers rotated by $30^\circ$ and $60^\circ$, respectively, as shown in Fig.~\ref{fig:dod 12-fold 3 layers}. More generally, the challenge of defining suitable coincidence windows (a) for different subsets of layers, when more than two layers are present; or (b) for different subsets of potential vertices within the coincidence set, as in the hexagonal Fibonacci tiling; as well as (c) using low-symmetry coincidence windows that break the full symmetry of the multilayer; will all require further exploration. 

Figure~\ref{fig:dod tilings 2 layers} illustrates some of this complexity. The tiling in Fig.~\ref{fig:dod 12-fold 3 layers} is generated from three square layers---red, blue, and green---whereby each vertex arises from the simultaneous near coincidence of points from all three layers, producing a dodecagonal tiling. In contrast, when only two layers are used---disregarding the green one, as shown in Fig.~\ref{fig:dod 4-fold 2 layers}---the tiling vertices become slightly displaced, producing a pseudo-dodecagonal tiling with only tetragonal symmetry. This can be seen by the appearance of distorted rhombs, with two distinct short-diagonal lengths, highlighted in purple and red. In Fourier space, this symmetry breaking manifests in a reduction in the intensity of some of the Bragg peaks, revealing the underlying 4-fold symmetry.

Finally, we plan to investigate bilayer and trilayer systems at very small twist angles, where the moiré patterns become exceptionally large. Such constructions may yield quasiperiodic tilings appropriate for modeling bilayer and trilayer graphene---systems of great current interest---and provide insight into the interplay between local near coincidences and long-range aperiodicity on experimentally relevant length scales.

\section*{Acknowledgments}

The authors are grateful to many colleagues, especially to Marianne Quiquandon and Denis Gratias, for their engaging comments during the  \emph{$15^{th}$ International Conference on Quasicrystals} (2023), where initial ideas about this work were first presented~\cite{Ochana23}, and later during the conference on \emph{Algebra, Analysis, and Aperiodic Order} (2024), dedicated to Michael Baake and Franz Gähler, on the occasion of their 65th birthdays. RL thanks Aviram Uri and Sergio de la Barrera for posing the challenge of assigning a proper quasiperiodic tiling to the trilayer graphene system, studied in Ref.~\cite{Uri23}, which we intend to address in a future publication. This research is funded by the Israel Science Foundation (ISF), through grant number 1259/22.

%

\appendix

\begin{table*}[hb]
\scriptsize
\centering
\renewcommand{\arraystretch}{1.2}
\begin{tabular}{|l|l|l|l|l|l|}
\hline
\hline
Family & Bilayer type & Window parameters (shape, rotation) & Edge length & Threshold & Figure \\
\hline
\hline
\multirow{2}{*}{\makecell{Octagonal \\(45$^\circ$ twist)}} 
  & \multirow{2}{*}{2 square lattices}
    & circle, \circwt & 1.2, 2.23, 2.9  & 0.224 & \ref{fig:square-kite-trapezoid} \\
    & & octagon, 0$^\circ$, \octagont[22.5]{} & 2.91 & 0.224 & \ref{fig:square-rhomb} \\

\hline
\multirow{10}{*}{\makecell{Dodecagonal \\(30$^\circ$ twist)}} 
  & \multirow{6}{*}{2 triangular lattices}
    & dodecagon, 15$^\circ$, \dodecagont[0]{}& \multirow{7}{*}{1.86} & 0.268 & \ref{fig:dod a} \\
  & & dodecagon, 0$^\circ$, \dodecagont[15]{} & & 0.299 & \ref{fig:dod b} \\
  & & union of 2 hexagons, $\pm15^\circ$, \hexagont[15]{} $\cup$ \hexagont[-15]{} & & 0.299 & \ref{fig:dod c} \\
  & & circle, \circwt & & 0.268 & \ref{fig:dod circ a} \\
  & & circle, \circwt & & 0.299 & \ref{fig:dod circ b} \\
  & & circle, \circwt & & 0.362 & \ref{fig:dod circ c} \\
  & & dodecagon, 0$^\circ$, \dodecagont[15]{} & & 0.374 & --- \\
\cline{2-6}
  & \multirow{2}{*}{3 square lattices}
    & dodecagon, 0$^\circ$, \dodecagont[15]{} & \multirow{2}{*}{2.15} & 0.29 & \ref{fig:dod 12-fold 3 layers} \\
  & & circle, \circwt & & 0.29 & --- \\
\hline
\multirow{2}{*}{\makecell{Fibonacci \\ ($\tau$-scaled)}} 
  & 2 square lattices & square, 0$^\circ$, \squarewt[45]{}  & 1.8, 3.11 & 0.437 & \ref{fig:square b} \\
  & 2 honeycomb tilings & circle, \circwt & 3.11, 4.92 & 0.383 & \ref{fig:hexagonal b} \\
\hline
\end{tabular}
\caption{Parameters for the quasiperiodic tilings generated by the near-coincidence method. 
The Table is grouped by family: octagonal tilings from 45$^\circ$ twisted bilayers, dodecagonal tilings from 30$^\circ$ twisted bilayers (and trilayers), and Fibonacci tilings from $\tau$-scaled bilayers. 
In each case we indicate the number and type of individual layers, the coincidence window (illustrated by a small symbol), the edge length, the threshold parameter, and a reference to the figure showing the tiling.}
\label{tab:tiling parameters}
\end{table*}


\section{Parameter values for producing the tilings displayed in this article} 
\label{sec:app-data}
Table~\ref{tab:tiling parameters} provides parameter values for different quasiperiodic tilings obtained by the near-coincidence method. 
For each bilayer type, we list the shape of the coincidence window, the edge length, and the threshold parameter. These tilings, illustrated throughout the paper, exemplify the versatility of the near-coincidence approach in reproducing well-known quasiperiodic tilings and generating new ones.

\section{An interactive web application of the near-coincidence method}
\label{sec:app-fib}
An interactive HTML demonstration of the near-coincidence method, developed with the assistance of the Gemini model~\cite{gemini23}, is available at:
\url{https://meshyochana.github.io/near_coincidence/}

Its interface is organized through a scrollable panel on the left side of the webpage, providing several controls: global settings for the displayed canvas and points; interaction-layer options, defining the threshold, the coincidence window, and the edge lengths; and layer-specific settings for parameters such as twist angle, scaling factor, and shift. The application is set for using the parameters given in Table~\ref{tab:tiling parameters}. Please cite the current article if using the application.

\end{document}